\documentclass[fleqn,10pt]{olplainarticle}
\usepackage{timet}
\usepackage{amsmath}
\usepackage{amsfonts}
\usepackage{tabularx}
\usepackage{graphicx}
\usepackage{physics}
\usepackage{url}
\usepackage{fancyhdr}
\usepackage[finalnew]{trackchanges} % inline for draft: for better track changes. finalnew option will compile

\author[Wouter Deleersnyder]
{Wouter Deleersnyder$^{1,2}$,David Dudal$^{1,3}$, Thomas Hermans$^{2}$,\\
	$^{1}$KU Leuven Campus Kortrijk - KULAK, Department of Physics, Etienne Sabbelaan 53, 8500 Kortrijk, Belgium.\\  E-mail: {wouter.deleersnyder@kuleuven.be} \\
	$^{2}$Ghent University, Department of Geology, Krijgslaan 281 - S8, 9000 Gent, Belgium \\
	$^{3}$Ghent University, Department of Physics and Astronomy, Ghent, Krijgslaan 281 - S9, 9000 Gent, Belgium \\
}
\date{Received 2022 XXXX XX; in original form 2022 XXXX XX}

\title{Novel Airborne EM Image Appraisal Tool for Imperfect Forward Modelling}

\author{Wouter Deleersnyder$^{1,2}$,  David Dudal$^{1,3}$, Thomas Hermans$^{2}$,\\
	$^{1}$KU Leuven Campus Kortrijk - KULAK, Department of Physics, Etienne Sabbelaan 53, 8500 Kortrijk, Belgium.\\  E-mail: {wouter.deleersnyder@kuleuven.be} \\
	$^{2}$Ghent University, Department of Geology, Krijgslaan 281 - S8, 9000 Gent, Belgium \\
	$^{3}$Ghent University, Department of Physics and Astronomy, Ghent, Krijgslaan 281 - S9, 9000 Gent, Belgium \\}

\begin{abstract}
Full 3D inversion of time-domain Airborne ElectroMagnetic (AEM) data requires specialists' expertise and a tremendous amount of computational resources, not readily available to everyone. Consequently, quasi-2D/3D inversion methods are prevailing, using a much faster but approximate (1D) forward model. We propose an appraisal tool that indicates zones in the inversion model that are not in agreement with the multidimensional data and therefore, should not be interpreted quantitatively. The image appraisal relies on multidimensional forward modelling to compute a so-called normalized gradient. Large values in that gradient indicate model parameters that do not fit the true multidimensionality of the observed data well and should not be interpreted quantitatively. An alternative approach is proposed to account for imperfect forward modelling, such that the appraisal tool is computationally inexpensive. The method is demonstrated on an AEM survey in a salinization context, revealing possible problematic zones in the estimated fresh-saltwater interface.
\end{abstract}

\begin{document}

\flushbottom

\maketitle
\textbf{\color{red} Submitted to Remote Sensing}
\thispagestyle{empty}

		\section{Introduction}

The Airborne ElectroMagnetic induction (AEM) method is a practical tool to map near-surface geological features over large areas, as electromagnetic induction methods are sensitive to the bulk resistivity. It is increasingly used for mineral exploration \citep{macnae2007developments}, hydrogeological mapping \citep{mikucki2015deep, podgorski2013processing}, saltwater intrusion \citep{goebel2019mapping, siemon2019automatic, deleersnyder2022flexible} and contamination \citep{pfaffhuber2017delineating}. AEM methods will become more and more important for the challenges in the future, e.g., as an important investigation method for groundwater management. It is the only viable approach to providing hydrogeological mappings on a large scale. Among the geophysical EM methods, the advancement of the AEM within the last two decades method was eminent.  While the AEM systems have massively advanced \citep{auken2017review}, the data interpretation process and the related computational burden remains a main impediment. Full 3D inversion is an active research area \citep{engebretsen2022accelerated,heagy2017framework, cai2017finite, yin20163d, ansari2017gauged,borner2015three, cox20103d}. It requires specialists' expertise and a tremendous amount of computational resources, not readily available to everyone. Consequently, quasi-2D and quasi-3D inversion methods are prevailing, using a much faster but approximate (1D) forward model. While using a 1D forward model is valid for slowly varying lateral variations, the hypothesis is not always valid. The question remains whether the obtained inversion results are reliable and can be interpreted quantitatively. In this work, we do not want to dissuade the use of 1D forward models for AEM interpretation. Rather, we argue that an additional step after each inversion with an approximate forward model should be added using an image appraisal tool, to verify that no erroneous interpretation has occurred as a result of the approximate \add{forward} model. \add{The tool indicates uncertain areas in the recovered model, which should be interpreted with extra care or should be reinterpreted using a full 3D inversion. In the latter case, this computationally demanding 3D inversion must, fortunately, only be performed on a subset of the original dataset.} \\

Appraisal tools usually address resolution issues. They are commonly used in electrical resistivity tomography \citep{oldenburg1999estimating, binley2005dc, caterina2013comparison}, with e.g. \citet{paepen2022effect} showing an application directly functional in a saltwater intrusion context. Specifically for EM, \citet{alumbaugh2000image} provide an appraisal tool based on the resolution matrix which provides insight on the resolution and accuracy of the recovered images. \citet{christiansen2003quantitative} provide a quantitative appraisal for AEM by adding a comparison to ground-based data. The method relies on 1D forward modelling and does not account for multidimensionality effects. \\

To overcome the latter shortcomings, we propose a novel appraisal tool that can detect wrongly fitting multidimensional data i.e., zones in the inversion model that are not in agreement with the multidimensional (2D/3D) forward model and therefore, should not be interpreted in a quantitative fashion. To our knowledge, such a tool has never been presented in the scientific literature. As generating multidimensional data in a time-domain AEM setting can be challenging, a successful, alternative approach is presented to function with imperfect forward 2/3D modelling. This allows for more accessible, computationally tractable computations on coarser discretizations on a single laptop with only a fraction of the required resources for perfect modelling.

\section{Method}
\subsection{Three Types of Forward Modelling}
\begin{figure}
	\noindent
	\begin{tabularx}{\textwidth}{rXp{0.25cm}Xrr|l}
		
		& \bf \large \hspace{10mm} Input &~&\bf \large Forward Model & ~&\bf \large Output  &\bf \large Fidelity \\
		A.&\noindent\parbox[b]{\hsize}{\includegraphics[width=100pt]{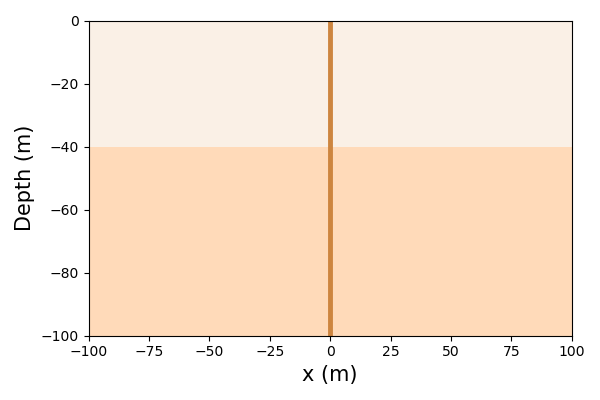} } &\large$\to$ &
		\includegraphics[width=100pt]{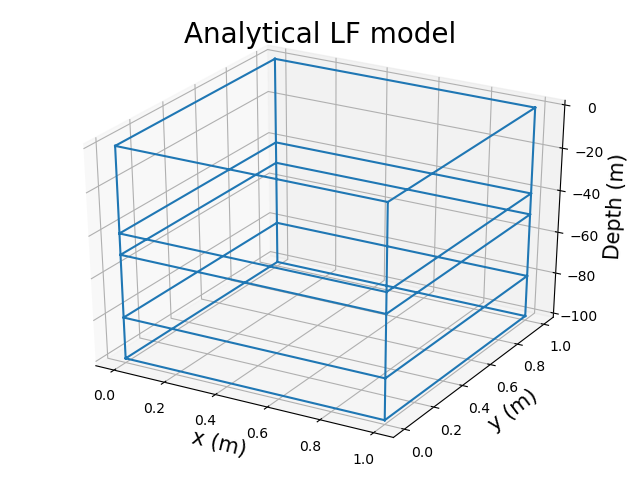} &\large=& $\mathcal{F}_{\text{1D}}(\vb{m})^* $ & Low (LF)\\
		
		B.& \includegraphics[width=100pt]{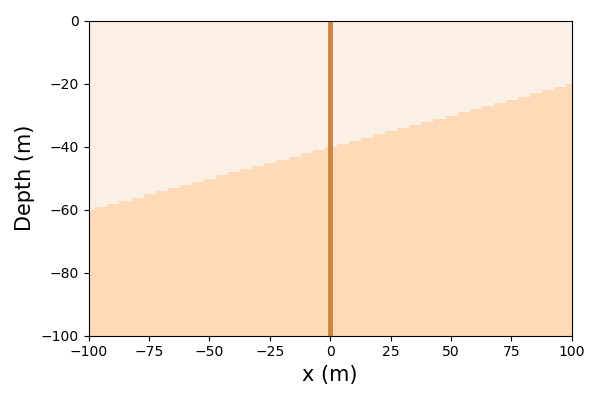}&\large$\to$  &\includegraphics[width=100pt]{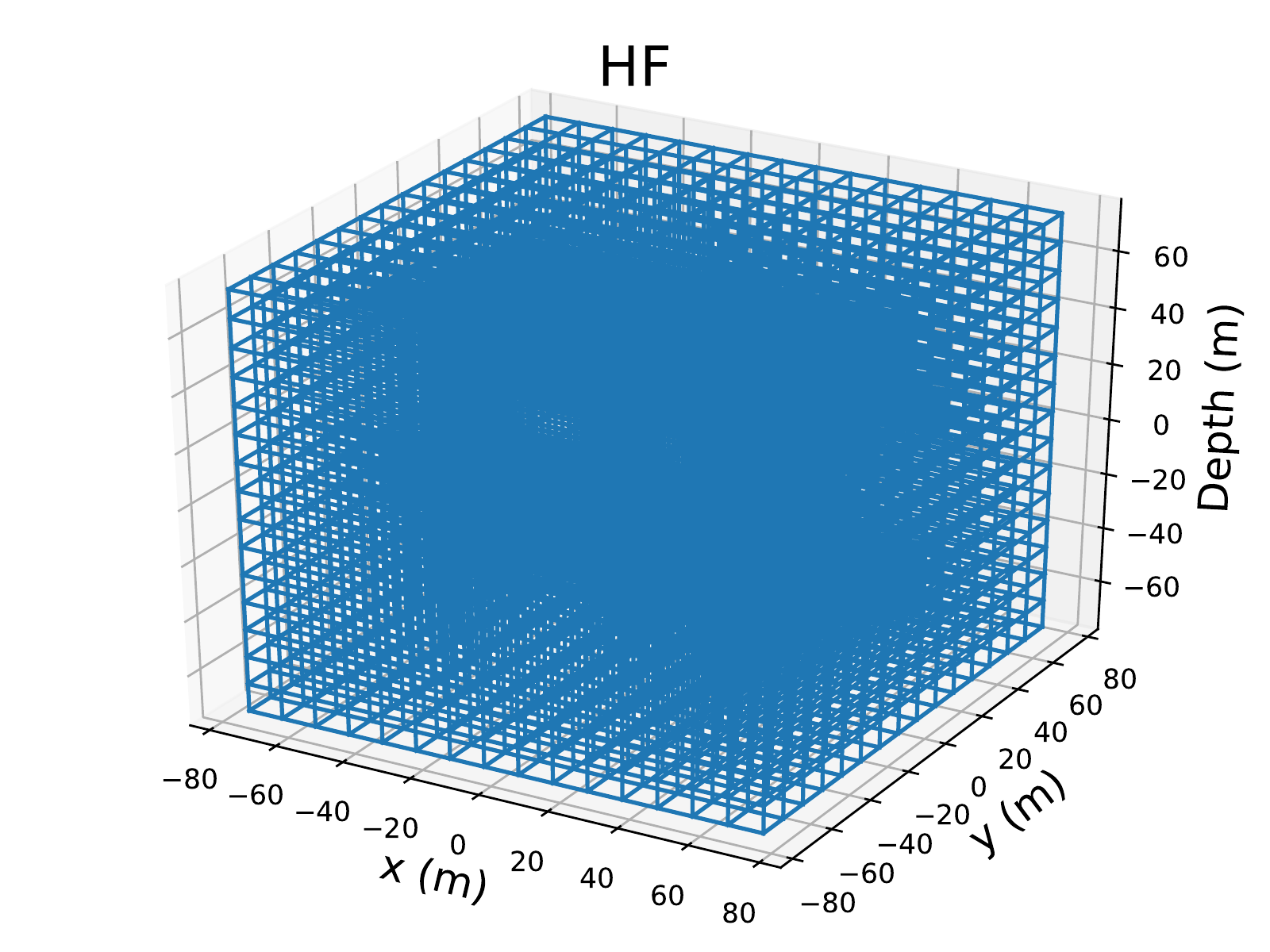} &\large=& $\mathcal{F}_{\text{2.5D}}  (\vb{m})$& High (HF)\\
		
		C. &\includegraphics[width=100pt]{fig/model_sub_1.png}&\large$\to$  &\includegraphics[width=100pt]{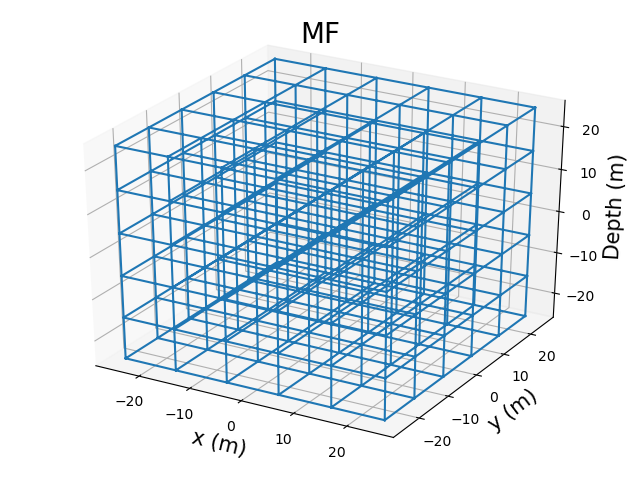} &\large=& $\mathcal{F}_{\text{2.5D}}  (\vb{m})$ & Medium (MF)\\
		
		D. &\includegraphics[width=100pt]{fig/model_constant_1.png}&\large$\to$  &\includegraphics[width=100pt]{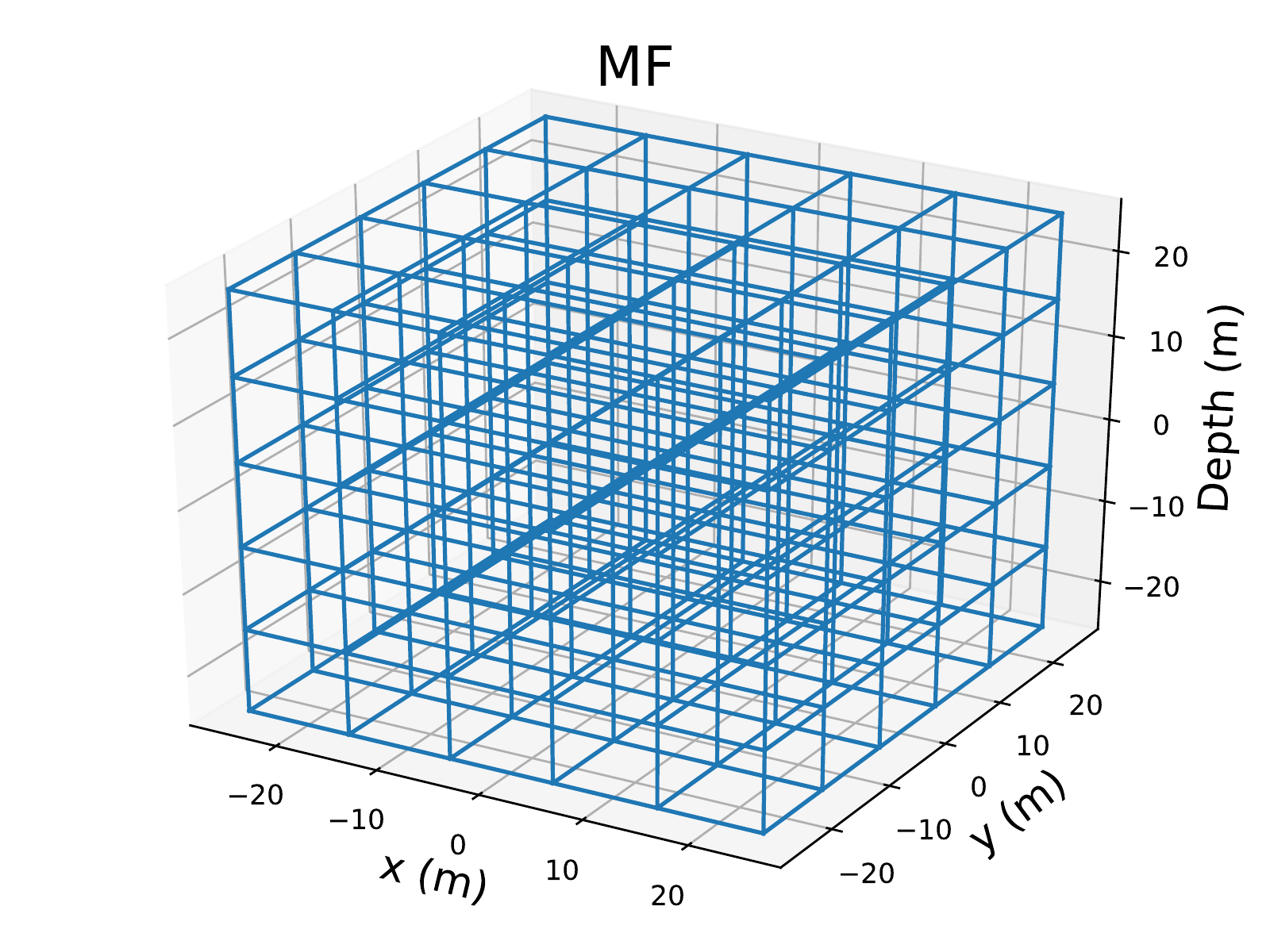} &\large=& $\mathcal{F}_{\text{2.5D}}  (\vb{m}^{\text{1D}})$& Medium  (MF)\\

	\end{tabularx}
	$^*$ In the 1D case, notation $\mathcal{F}_{\text{1D}}(\vb{m}) = \mathcal{F}_{\text{1D}} (\vb{m}^{\text{1D}}) $
	\caption{Conceptual visualization of the various forward data types used. The input is either a multidimensional subsurface model (B. and C.) or an 1D subsurface model (in a moving footprint approach) (A. and D.). The forward model is either a Low-Fidelity (LF) analytical forward model (with depths and EC as input) (see A.), a High-Fidelity (HF) forward simulation on an accurate mesh (B.) or a Medium- Fidelity (MF) forward simulation on an inaccurate (coarser) mesh (C. and D.). NOTE: the presented meshes are illustrative and are not the ones used for multidimensional modelling. Some details on the MF and HF mesh are described in Appendix \ref{sec:app}.}
	\label{tab:label} 
	\note{This figure is added w.r.t. the original submission.}
\end{figure}

The forward model describes the subsurface's response to a specific subsurface realization and a specific survey set-up. There are two main common approaches: The first is based on (semi-)analytical models that solve the (continuous) Maxwell equations for a one dimensional subsurface model, meaning that it assumes horizontal layers without lateral variations. An open-source Python implementation by \citet{werthmueller2017open} neatly implements such a forward model by \citet{hunziker2015electromagnetic} in a fast and reliable fashion. We refer to this model as the low-fidelity model (LF), as it cannot account for lateral variations in the subsurface model. The second approach is based on a discretization of the physics on a mesh. Those simulations mimic the full 3D soil response of the potentially non-1D subsurface and allow for multidimensional modelling.  In this work, the finite volume method from the open-source package SimPEG \citep{heagy2017framework} is used. With a suitable discretization of the geometry, an accurate magnetic field response can be obtained. In the case of perfect forward modelling, we refer to these simulations as the high-fidelity model (HF). However, numerical simulations are not always accurate and the term high-fidelity should be used with caution. If the accuracy of the simulations is limited due to the computational burden requiring the use of a coarse mesh, the response contains a modelling error. That modelling error is different in origin than the one introduced by only considering a one-dimensional subsurface and depends on the discretization of the user and subsurface model. We refer to this model as the medium-fidelity model (MF). \add{We visualized the various types of modelling in Figure }\ref{tab:label}.

\subsection{Quasi-2D Inversion}
In most geophysical inverse problems, the inversion model $\vb{m}$ consists of electrical conductivities (EC) and fits the observed data $\vb{d}^{\text{obs}}$ and is simple in Occam's sense \citep{constable1987occam}. This is accomplished by minimizing an objective function
\begin{linenomath*}
	\begin{equation}
		\phi(\vb{m}) = \phi_d(\vb{m}) + \beta\phi_m(\vb{m}),
	\end{equation}
\end{linenomath*}
where $\phi_d$ and $\phi_m$ are, respectively, the data and model misfit. $\beta$ is a regularization parameter which balances the relative importance of the two misfits.

In quasi-2D inversion, the data misfit
\begin{equation}
	\phi_d(\vb{m}) = \frac{1}{n} \left|\left|  \vb{W}_d\left( \vb{d}^{\text{obs}} - \mathcal{F}_{\text{1D}}(\vb{m}) \right) \right|\right|_2^2
\end{equation}
uses a 1D approximation for the forward (LF) model $\mathcal{F}_{\text{1D}}(\vb{m})$, which significantly reduces the computation time of the inversion procedure. The model misfit $\phi_m$ promotes smooth solutions \citep{tikhonov1943stability, constable1987occam}.\\

In this work, the regularization parameter $\beta$ is selected via the chi-squared criterion, meaning that an optimal inversion model fits the observed data to an  noise-weighted RMS\change{, where the noise is defined in the diagonal matrix $ \vb{W}_d$}{. The diagonal matrix $ \vb{W}_d$ contains the noise-levels, that is the reciprocals of the estimated noise standard deviation and the noise floor. The latter ensures that not too much weight is given to the last channels, as they are highly sensitive to measurement error due to their small absolute value (in this work, the noise floor is set to $10^{-13}$).} As we work with two forward models, we distinguish two RMS errors:
\begin{linenomath*}
	\begin{equation}
		\epsilon_{\text{1D}} =  \sqrt{\frac{1}{n} \left|\left|  \vb{W}_d\left( \vb{d}^{\text{obs}} - \mathcal{F}_{\text{1D}}(\vb{m}) \right) \right|\right|_2^2}, \qquad 	\epsilon_{\text{2.5D}} =  \sqrt{\frac{1}{n} \left|\left|  \vb{W}_d\left( \vb{d}^{\text{obs}} - \mathcal{F}_{\text{2.5D}}(\vb{m}) \right) \right|\right|_2^2},
	\end{equation}
\end{linenomath*}
where $\mathcal{F}_{\text{2.5D}}$ refers to \add{either} the high- or medium-fidelity \add{forward} model which \change{allows}{better allow} for 2D variations on a 3D mesh. The approach could be extended without loss of generality to a full 3D forward model\add{, meaning that 3D variations are modelled on a 3D mesh}.

\subsection{Normalized Gradient}
The multidimensional sensitivity matrix or Jacobian $\vb{J}\left( = \frac{\partial \vb{d}}{\partial \vb{m}}\right)$  is required to map poorly fitting data points to specific areas of the inversion model. A high sensitivity value signifies that a change of this parameter influences the predicted data strongly. We propose to compute the sensitivity matrix $\vb{J}_i$ on a coarse 3D mesh with a strongly reduced mesh size and number of cells, optimized for computing the response for one sounding and allowing for a computation on a single laptop/\add{desktop}. This strategy combines the moving footprint approach \citep{cox20103d} and that of \citet{zhang20213d}.\\

With the computed data, the normalized gradient $\tilde{\nabla} \phi_{d, 2.5D} $ is computed as follows:
\begin{linenomath*}
	\begin{equation}
		\tilde{\nabla} \phi_{d, 2.5D} = \frac{\sum_i |\vb{J}_i^{\text{T}} \vb{W}_d\left( \vb{d}_i^{\text{obs}} - \mathcal{F}_{\text{2.5D}}(\vb{m})_i \right)| }{\sum_i |  \vb{J}_i^{\text{T}} | },
		\label{eq:gradient}
	\end{equation}
\end{linenomath*}
where the $i$ refers to a specific sounding. Note that this computation may involve an interpolation step to map all data to, for example, the discretization of the recovered model. We are mainly interested in the relative importance of each zone in the inversion model, hence the denominator. The normalised gradient $\tilde{\nabla} \phi_{d, 2.5D} $ gives an indication for which model parameters would change in a full 2D inversion, meaning that those model parameters do not fit the data well with a multidimensional forward model. Put differently, model parameters that would not rapidly change are likely to be fitting the data well and can be interpreted quantitatively. 

\subsection{Accounting for Imperfect Modelling}
The image appraisal method does not need to be used with an expensive, exact high-fidelity forward model $\mathcal{F}_{\text{2.5D}}$. If a medium fidelity model is used, the predicted data $\mathcal{F}_{\text{2.5D}}(\vb{m})$ for the recovered model will not be fitting the observed data $\vb{d}^{\text{obs}}$ as well as the 1D forward model $\mathcal{F}_{\text{1D}}$, which we refer to as the forced modelling error approach. \add{However, our method will still allow to identify in which zone of the model multi-dimensional effects are significant.}\\

%\textit{1. Robust error selection method. } Due to the modelling error, many individual ($^i$) data points $\epsilon_{\text{2.5D}}^i$ can have errors greater than one. Only the largest errors should be selected. The quantile transform transforms the error distribution $\epsilon_{\text{2.5D}}^i$  to a Gaussian distribution, in which the standard deviation $\sigma$ is calculated. Only errors greater than 3/2$\sigma$ are selected and contribute in the numerator of normalized gradient in eq. \eqref{eq:gradient}. The modelling error depends on the gate time  (usually a certain discretization/mesh has its limitation in a certain range), therefore the quantile transform is applied to the distribution of the errors per gate time. This is further demonstrated in Section \ref{sec:robust}.

In this approach the model parameters right below the sounding location are fed as 1D subsurface model to the imperfect 2.5D medium-fidelity forward model $\mathcal{F}_{\text{2.5D}}$. As a result, we get less accurate data than with the 1D forward model $\mathcal{F}_{\text{1D}}$, but with similar modelling errors per gate time. The normalized gradient \eqref{eq:gradient} is otherwise identical, but $\vb{d}^{\text{obs}}$  is replaced by $\mathcal{F}_{\text{2.5D}}(\vb{m}^\text{1D})$, where $\vb{m}^\text{1D}$ represents the 1D subsurface model per sounding, which is generated from the recovered model. This approach is demonstrated in Section \ref{sec:forced}. 

%
%\subsection{Proposed Workflow for Low-Cost Image Appraisal}
%1. After each quasi 2D or 3D inversion result, verify the RMS error with a high-fidelity forward model 	$\epsilon_{\text{2.5D}}$.\\
%2. If the reliable RMS error is problematic, compute the approximated sensitivity matrix for each data point. Compute the normalized gradient $\tilde{\nabla} \phi_{d, 2.5D}$ to determine the problematic zones.\\
%3. Look at how the normalized sensitivity would alter the inversion model. We avoid the trap of interpreting this information quantitatively.\\
%4. If specific features cannot be interpreted quantitatively, crucial to the research objectives, then a full 2D/3D inversion is warranted (if sufficient computational resources are available).

\section{Results}
In this section, we apply our proposed methodology on a synthetic model and a real field data case within a saltwater intrusion context \citep{vlaanderentopsoil}, both with the time-domain AEM data from a dual moment (LM+HM) SkyTEM instrument \citep{sorense2004skytem}.
\subsection{Synthetic Model}
\begin{figure}[!htb]
	\footnotesize 
	A. \hspace{0.49\textwidth} B.\\
	~\includegraphics[width=0.49\textwidth]{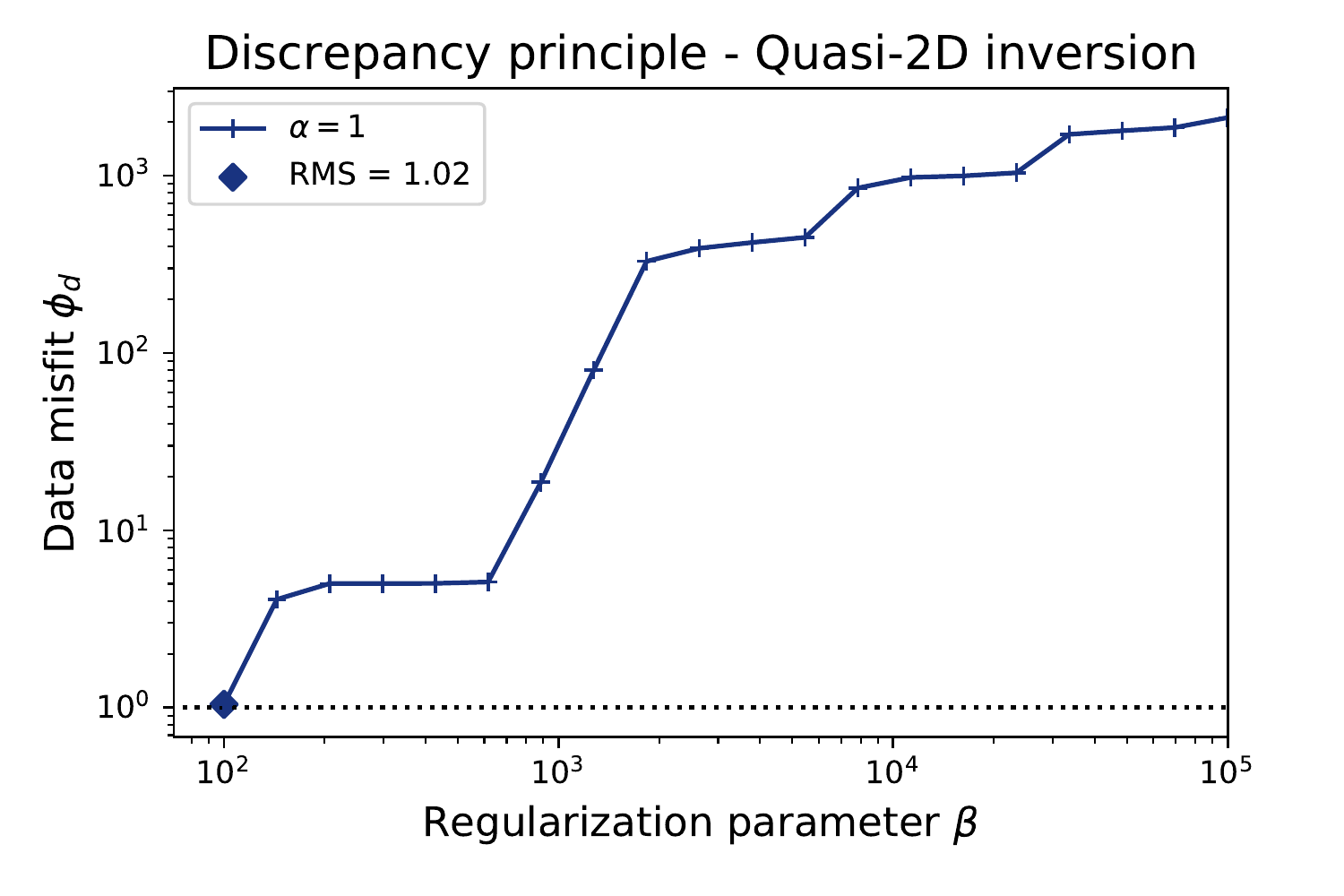} 
	\includegraphics[width=0.49\textwidth]{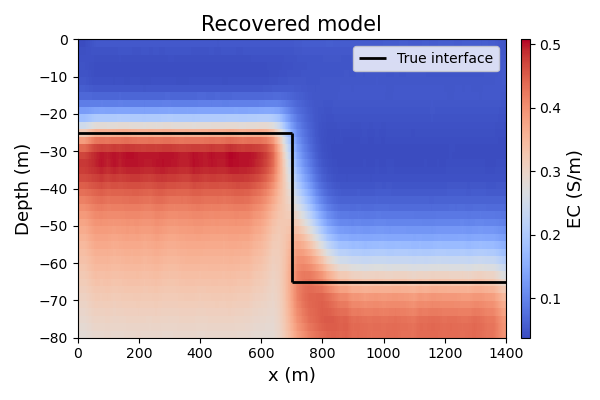} 
	
	C. \hspace{0.49\textwidth} D.\\
	\includegraphics[width=0.49\textwidth]{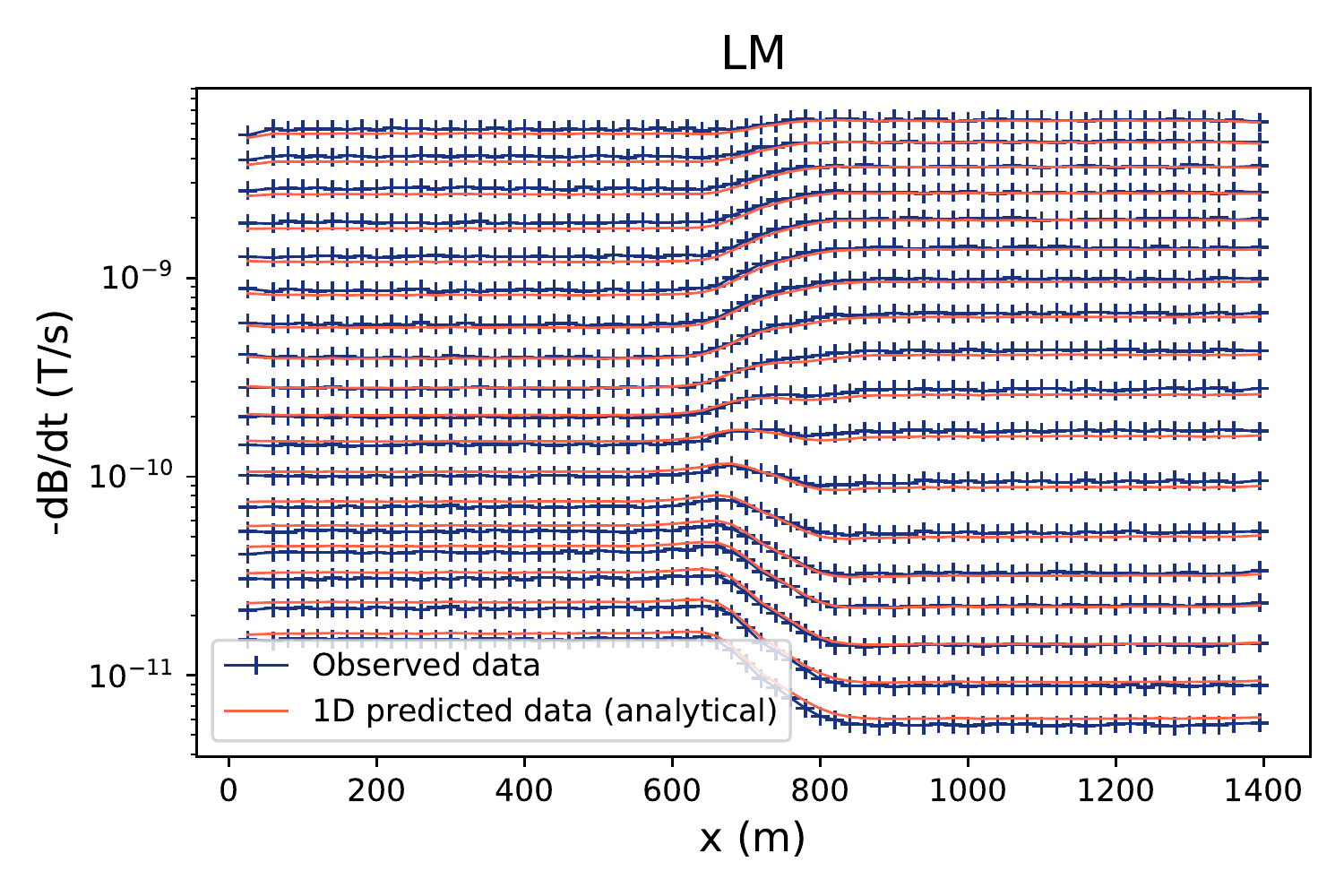} 
	\includegraphics[width=0.49\textwidth]{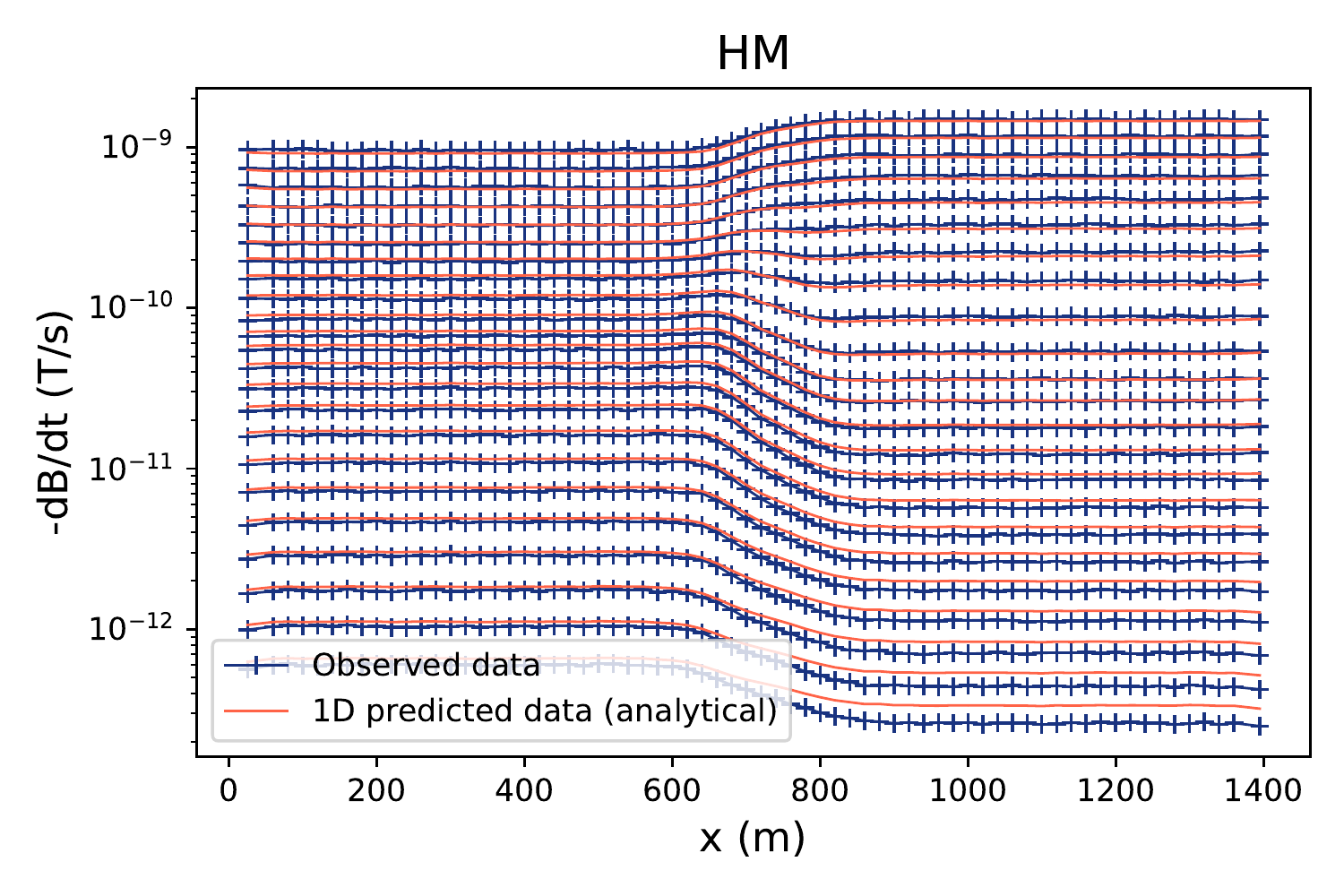} 
	\caption{\add{Quasi-2D inversion results. }A. Discrepancy principle \add{for the selection of the optimal regularization parameter $\beta$}. B. Recovered model. C.-D. Observed and predicted data (time-domain electromagnetic induction data) with a 1D analytical forward model $\mathcal{F}_{\text{1D}}$.}
	\label{fig:recovered_model}
\end{figure}

Any appraisal method for multidimensional effects should indicate areas in the recovered model that potentially do not fit the true multidimensionality of the observed data well. \change{On the contrary,}{Simultaneously,} it should not indicate areas in the recovered model where there appear to be no issues with data interpretation. For field data, where the true subsurface parameters are unknown, our proposed method is difficult to validate. Therefore, a simple subsurface model was created to demonstrate and verify the performance of our method, before applying it to  real field data in Section \ref{sec:field}.\\

Consider the recovered model in Figure \ref{fig:recovered_model}B, which consists of two layers of 0.05 S/m and 0.5 S/m, respectively. The model is selected based on the discrepancy principle (Figure \ref{fig:recovered_model}A)\add{, which states that the optimal value for the regularization parameter corresponds to the case in which the data fits up to the noise level, i.e., $\phi_d \approx 1$ }\citep{hansen2010discrete}. The recovered model has a noise-weighted error $\epsilon_{\text{1D}}$ of 1.02, while with the multidimensional data $\epsilon_{\text{2.5D}}$ it is 1.53. \add{The latter indicates that the model does not fit the data to its noise level when a HF, multi-dimensional forward model is used.} The predicted data with the low-fidelity model and the observed data points are presented in Figure \ref{fig:recovered_model}C-D\add{, where the slight discrepancy in the late HM time channels are ascribed to the noise floor}.  In the recovered model, the interface between both layers changes abruptly at $x = 700$. Near the interface, the recovered model suggests a dipping layer and the interpretation can be erroneous without taking into account the use of the 1D forward model for the generation of the recovered model. 

\subsubsection{Perfect Modelling Approach}
\label{sec:perfect}
\begin{figure}
	
	\footnotesize 	A.  \hspace{0.5\textwidth} 	\footnotesize 	 B.\\
	\includegraphics[width=1\textwidth]{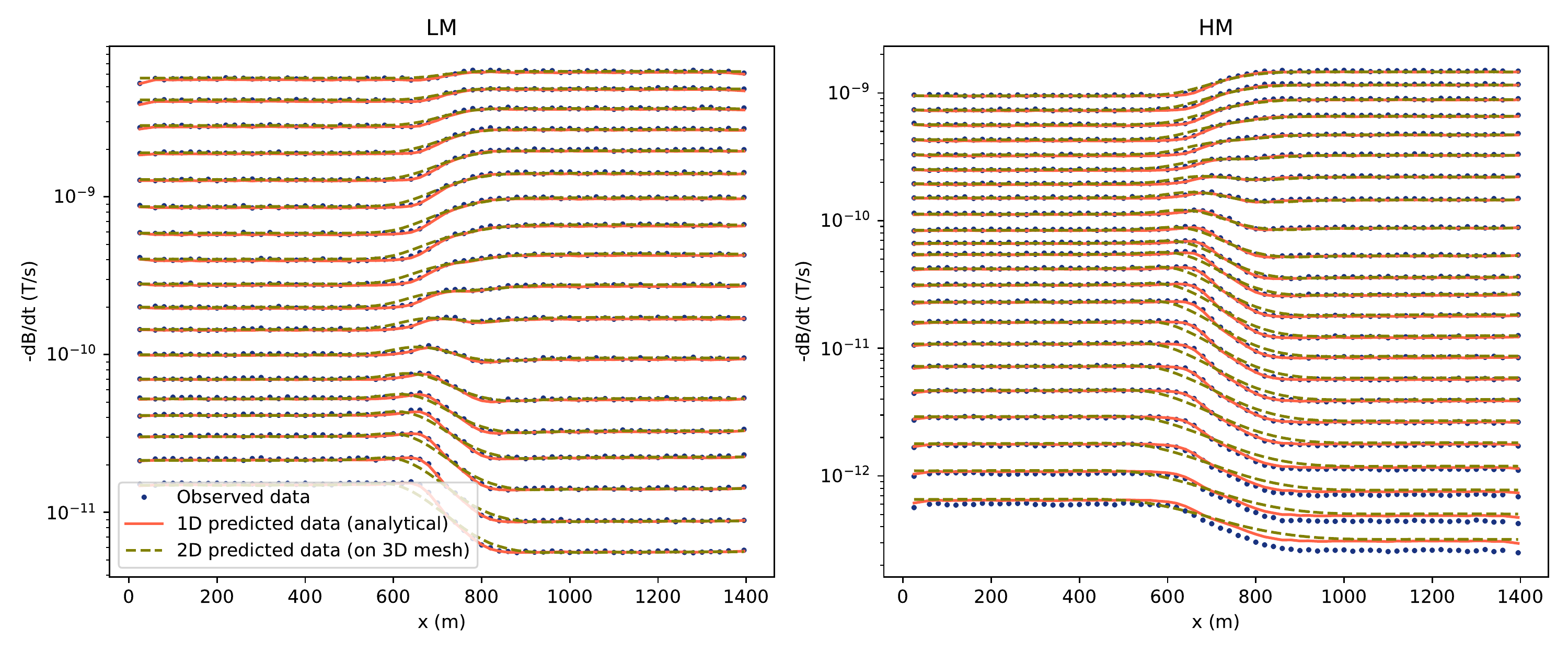} 
	\footnotesize 	C.  \hspace{0.5\textwidth} 	\footnotesize 	 D.\\
	\includegraphics[width=0.49\textwidth]{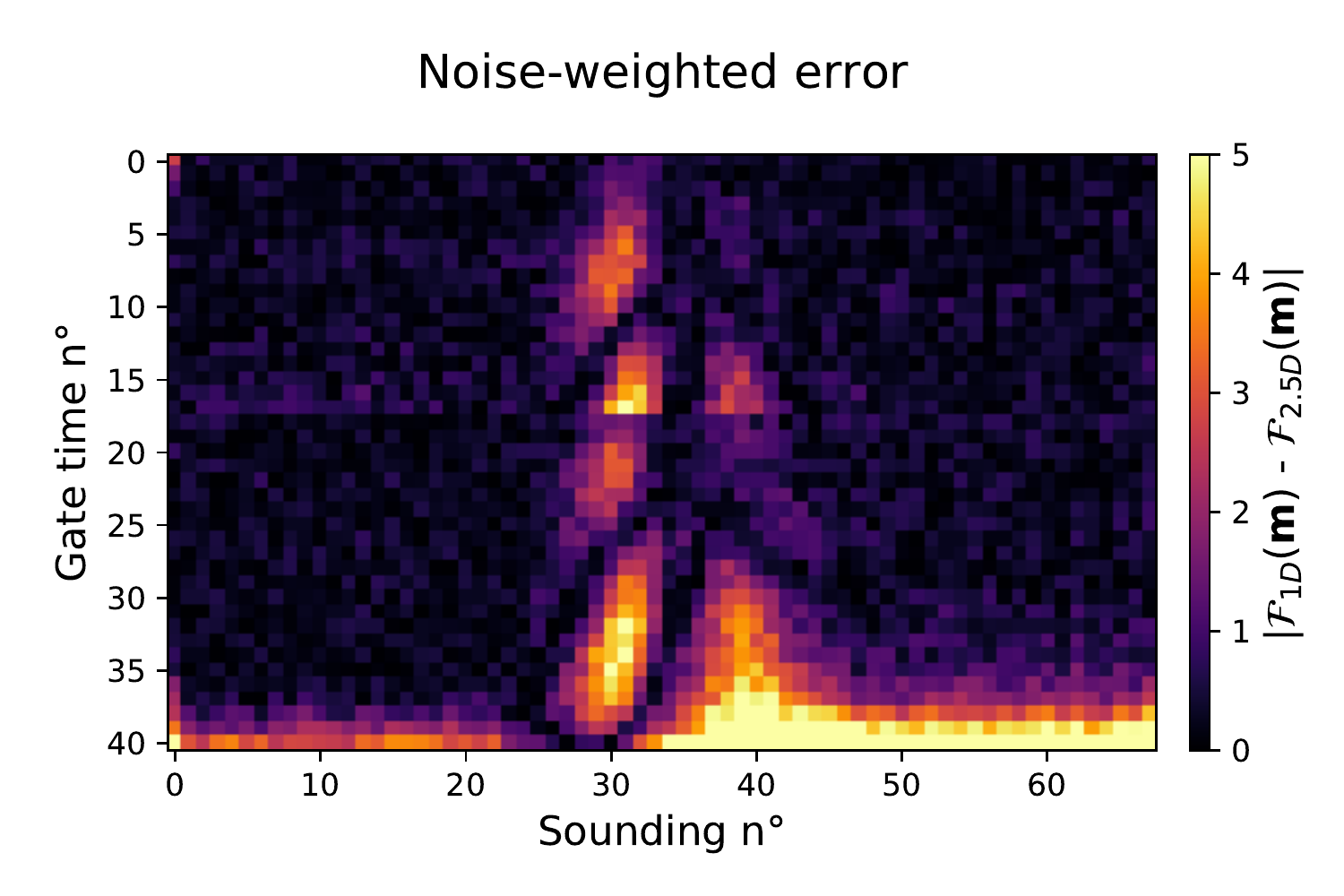} 
	\includegraphics[width=0.49\textwidth]{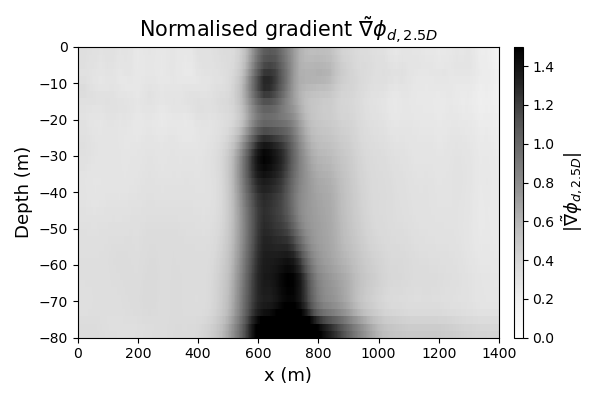} 
	\footnotesize 	E.  \hspace{0.5\textwidth} 	\footnotesize 	 F.\\
	\includegraphics[width=1\textwidth]{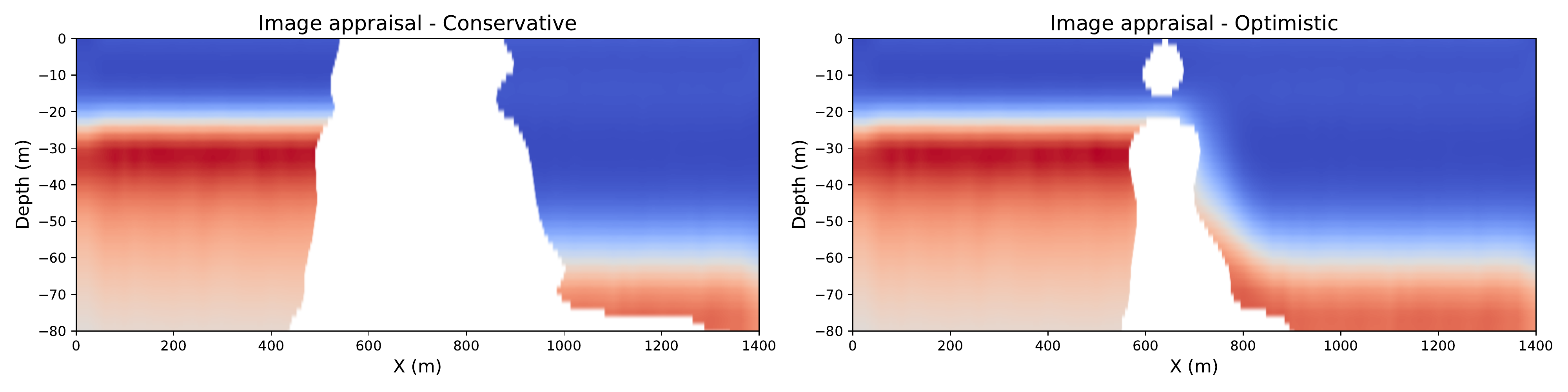} 
	\caption{Image appraisal with \change{perfect}{\emph{perfect}} forward modelling.
		\add{A.-B. A discrepancy mainly near the $x = 700$ m interface. C. The individual noise-weighted errors. The first 18 gate times correspond to the LM and the 23 last gate times to the HM. D. The normalised gradient indicates areas near the $x = 700$ m interface. E.-F. Image appraisal outcomes clearly indicate problematic areas in the region near the 'step'. }}
	\label{fig:synthetic_perfect}
\end{figure}

The first approach compares the predicted multidimensional data ($\mathcal{F}_{\text{2.5D}}$) with the observed data $ \vb{d}^{\text{obs}}$. In Figure \ref{fig:synthetic_perfect}A and \ref{fig:synthetic_perfect}B, the green dashed line from the predicted data $\mathcal{F}_{\text{2.5D}}(\vb{m})$ clearly deviates between 600 m and 800 m from the observed data points. The individual noise-weighted errors are also shown in Figure \ref{fig:synthetic_perfect}C.  For some soundings between n$^\circ$30 and n$^\circ$40 in Figure \ref{fig:synthetic_perfect}C, where $\vb{d}^{\text{obs}} = \mathcal{F}_{\text{2.5D}}(\vb{m})$, the inversion model \textit{appears} to fit the multidimensional data, it is an anticipated behaviour with this kind of subsurface model that cannot be determined objectively in a general fashion (for an unknown subsurface model). As the sensitivity functions overlap with neighbouring soundings, this poses no problems and this alleged perfect fit will also come into the scope of the appraisal method, which is evidently a plus.\\

The normalized gradient is shown in Figure \ref{fig:synthetic_perfect}D, which must be \change{viewed}{considered} together with the recovered model. The larger (darker) the normalised gradient, the more likely the interpretation in the area is incorrect. In Figure \ref{fig:synthetic_perfect}E-F the recovered model is shown where cells are left white if they are larger than a certain cut-off value, set by the user. In our case that is if they are larger than 20\% or 50\% of the maximum value in the normalized gradient (conservative, resp. optimistic case). The user can, as it were, scroll from \add{a} max normalized gradient to smaller values to get an indication of \remove{where }which areas are more/less likely to be correct.
\subsubsection{Imperfect Modelling - Approach with Forced Modelling Error}
\label{sec:forced}
The second approach uses a medium-fidelity model to compute the 1D model response. The model parameters right below the sounding location are used, as in the quasi-2D inversion scheme, and extended/projected to construct horizontally stratified layers to be used in the simulation. When using the same 2.5D forward model on the 1D model, a similar modelling error is expected than on the 2.5D data (while of course, we have used a faster and more accurate 1D forward model $\mathcal{F}_{\text{1D}}$ throughout the inversion). Then, the obtained data $\mathcal{F}_{\text{2.5D}}  (\vb{m}^{\text{1D}})$ can be compared to $\mathcal{F}_{\text{2.5D}}  (\vb{m})$  to get an idea of where the multidimensionality of the observed data is not well fitted. In eq~.~\eqref{eq:gradient}, $\vb{d}^{\text{obs}}$ is replaced with $\mathcal{F}_{\text{2.5D}}  (\vb{m}^{\text{1D}})$. A disadvantage of this method is that the observed data is no longer used. The advantage is that it is not required to work with (robust) selection methods to eliminate the modelling error (which can potentially fail for poorly designed meshes).\\

\begin{figure}
	\footnotesize 	A.  \hspace{0.5\textwidth} 	\footnotesize 	 B.\\
	\includegraphics[width=1\textwidth]{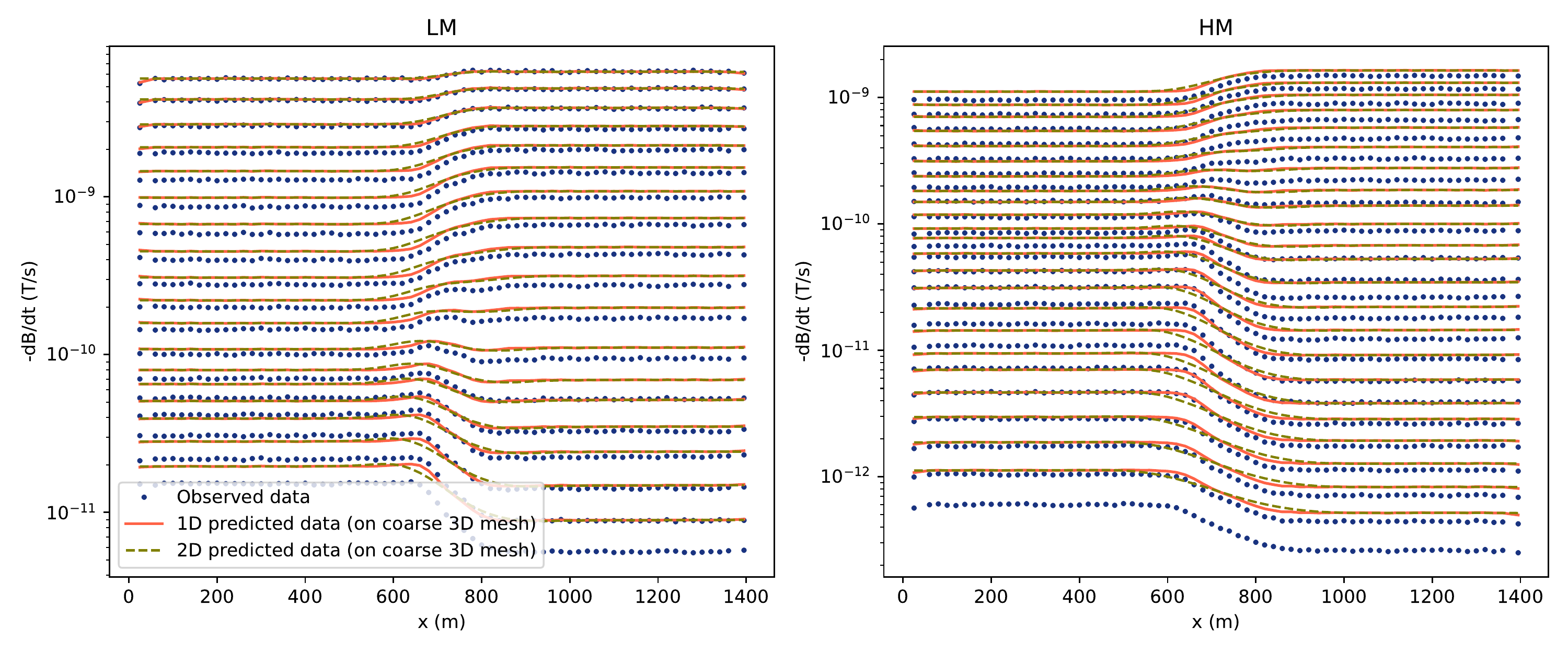} 
	\footnotesize 	C.  \hspace{0.5\textwidth} 	\footnotesize 	 D.\\
	\includegraphics[width=0.49\textwidth]{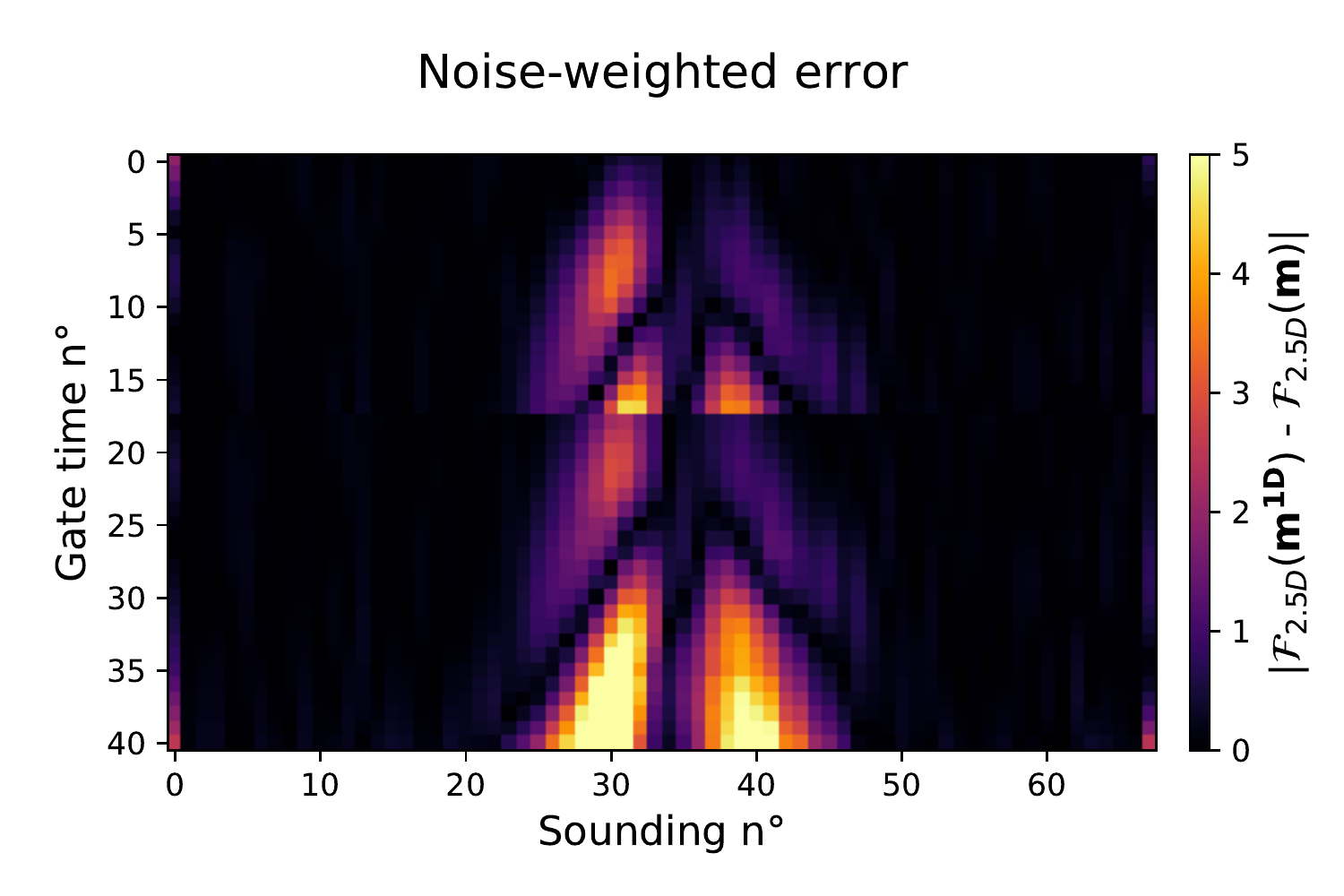} 
	\includegraphics[width=0.49\textwidth]{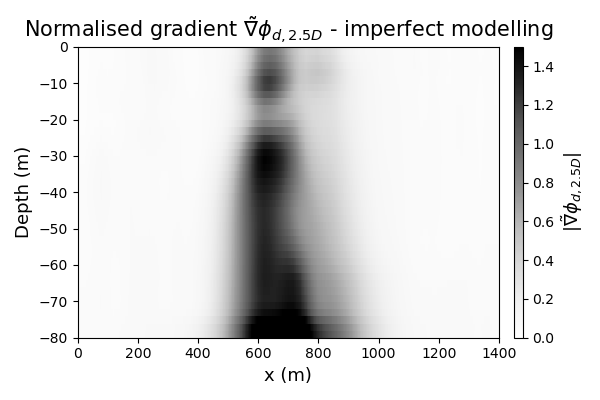} 
	\footnotesize 	E.  \hspace{0.5\textwidth} 	\footnotesize 	 F.\\
	\includegraphics[width=1\textwidth]{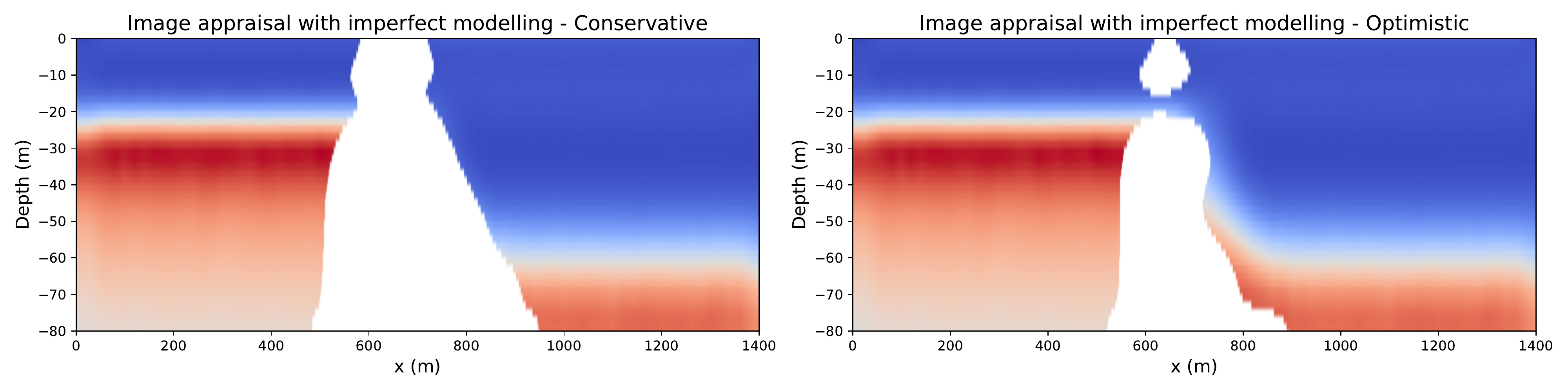} 
	
	\caption{Image appraisal with \textit{imperfect} forward modelling approach.
		%A.-B. Observed and predicted data with low and high-fidelity forward model, C. Noise-weighted errors. D. Normalised gradient. E.-F. Conservative and optimistic image appraisal results.
		\add{A.-B. A discrepancy mainly near the $x = 700$ m interface. C. The individual noise-weighted errors. The first 18 gate times correspond to the LM and the 23 last gate times to the HM.  D. The normalised gradient indicates areas near the $x = 700$ m interface. E.-F. Image appraisal outcomes clearly indicate problematic areas in the region near the 'step'. }
	}
	\label{fig:synthetic_imperfect}
\end{figure}

The procedure behind the image appraisal is analogous to the previous approach in Section \ref{sec:perfect}. Contrary to the previous method, additional data is generated with the 2.5D forward model  $\mathcal{F}_{\text{2.5D}}$ of the 1D version of the recovered version $\mathcal{F}_{\text{2.5D}}  (\vb{m}^{\text{1D}})$ . This data is presented in Figure \ref{fig:synthetic_imperfect}A and \ref{fig:synthetic_imperfect}B as the green dashed line. It is apparent that the predicted data \change{$\mathcal{F}_{\text{1D}}$}{$\mathcal{F}_{\text{2.5D}}  (\vb{m}^{\text{1D}})$ } and  $\mathcal{F}_{\text{2.5D}}(\vb{m})$ overlap at the horizontal parts of the recovered model, while a deviation is observed at the transition near $x$ = 700 m. The \change{$\mathcal{F}_{\text{1D}}$}{predicted} seems to follow the trend in the observed data well, while the exact value is quite different. \add{This should be ascribed to the multidimensionality of the step at  $x$ = 700 m, but the imperfect modelling of the MF forward model. More background is provided in Appendix }\ref{sec:app}\add{ and in Figure }\ref{fig:bad_sounding}\add{, more specifically.} The noise-weighted errors are no less than $\epsilon_{\text{1D}} = 14.6$ and $\epsilon_{\text{2.5D}} = 14.7$, signalling a significant modelling error. The absolute values of the noise-weighted errors with respect to the true observed data are shown in Figure \ref{fig:synthetic_imperfect}C. The errors around $x$ = 700 m are prominent. The normalized gradient is presented in Figure \ref{fig:synthetic_imperfect}B. It closely resembles the normalized gradient with perfect modelling. The image is less noisy, as we are no longer comparing with the observed data and thus the measurement error is lacking. The resulting image appraisal images are presented in Figure \ref{fig:synthetic_imperfect}C-D and are very similar to the results from Section \ref{sec:perfect}, thereby reconfirm the interpretation already made there.\\

To illustrate the reduction in the computational burden, for the data production for a single sounding on the precise mesh for perfect modelling in Section \ref{sec:perfect}, 1h and 20 minutes of computation time were required on 36 cores of an HPC infrastructure node (2 x 18-core Intel Xeon Gold 6140 (Skylake @ 2.3 GHz)) and required 150 GB of RAM. The simulations on the coarse mesh for imperfect modelling were performed in \add{just a few} seconds on a laptop with Apple's M1 chip and 10 cores and with negligible memory usage.

\subsection{Field Data Case}
\label{sec:field}
\begin{figure}
	\centering
	\includegraphics[width=0.5\textwidth]{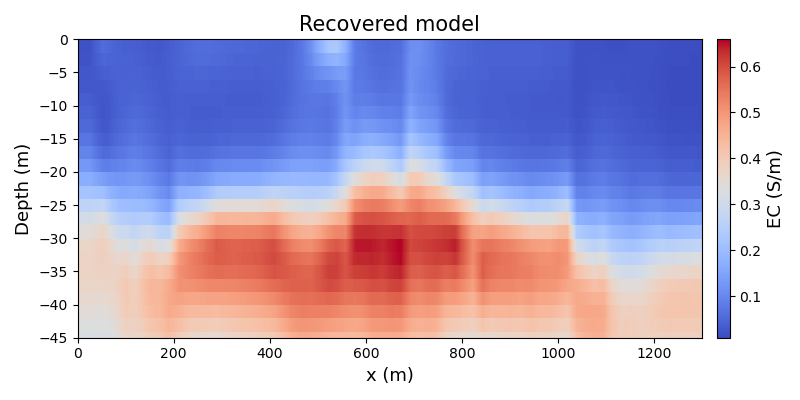} 
	\caption{Recovered model from real field data from \citet{vlaanderentopsoil}.}
	\label{fig:field_recovered}
\end{figure}

The image appraisal method is applied on time-domain AEM data with the setup described in \citet{vlaanderentopsoil}. The inversion model in Figure \ref{fig:field_recovered} is obtained via quasi-2D inversion. The obtained inversion model has $\epsilon_{\text{1D}}$ of 0.97, while 2.9 for $\epsilon_{\text{2.5D}}$. The result shows low values of Electrical Conductivity (EC) at shallow depths and high values of EC between 20 m and 50 m depth, which corresponds to a saltwater lens resting on a clay layer.  There is quite some lateral variation, so this is an interesting case for our appraisal tool.\\

\begin{figure}
	\footnotesize 	A.  \hspace{0.5\textwidth} 	\footnotesize 	 B.\\
	\includegraphics[width=0.49\textwidth]{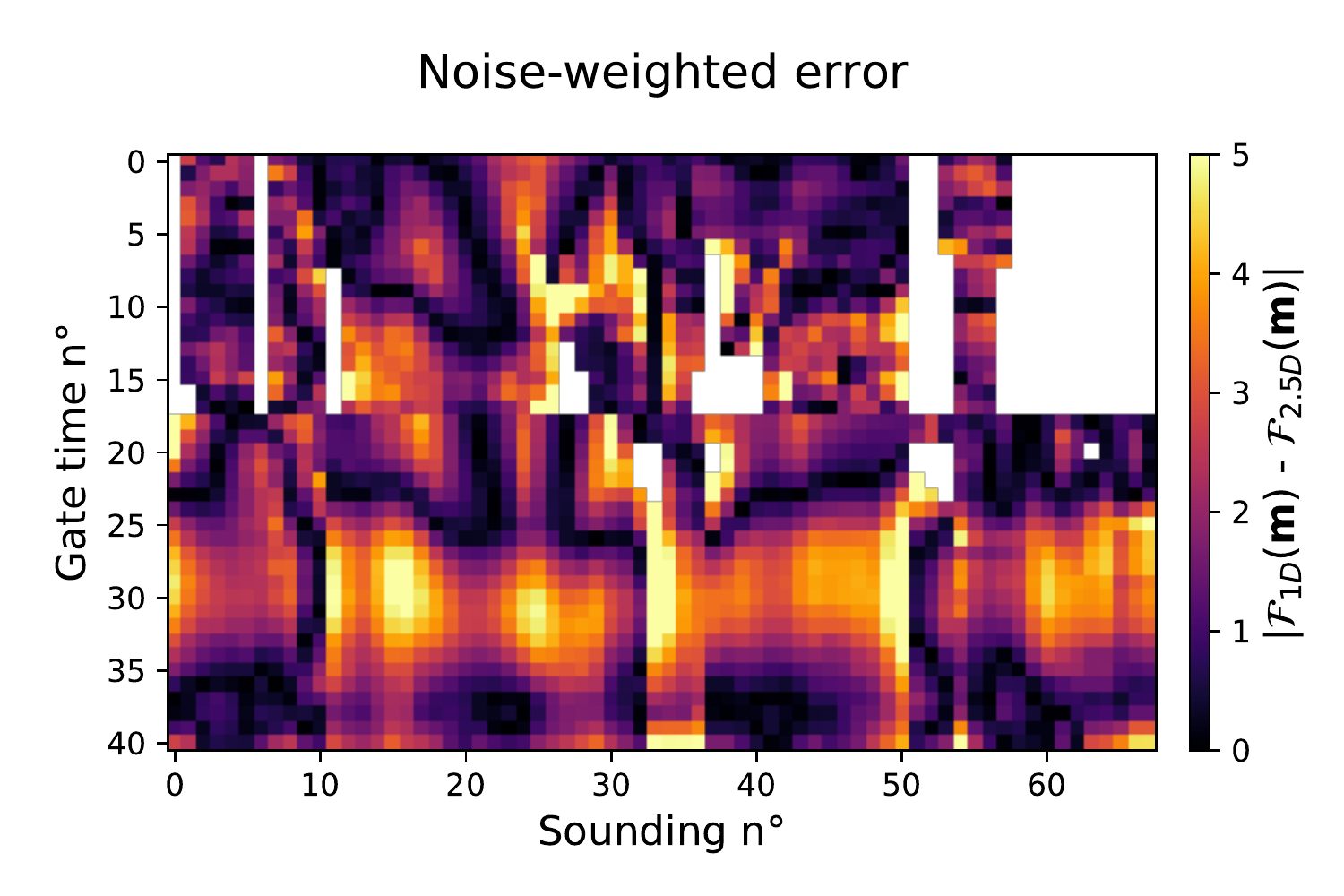} 
	\includegraphics[width=0.49\textwidth]{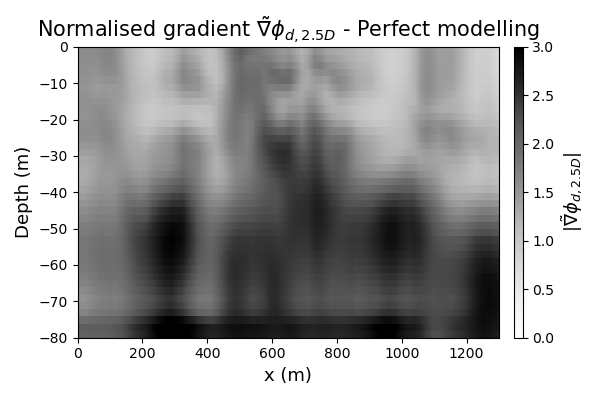} 
	\footnotesize 	C.  \hspace{0.5\textwidth} 	\footnotesize 	 D.\\
	\includegraphics[width=1\textwidth]{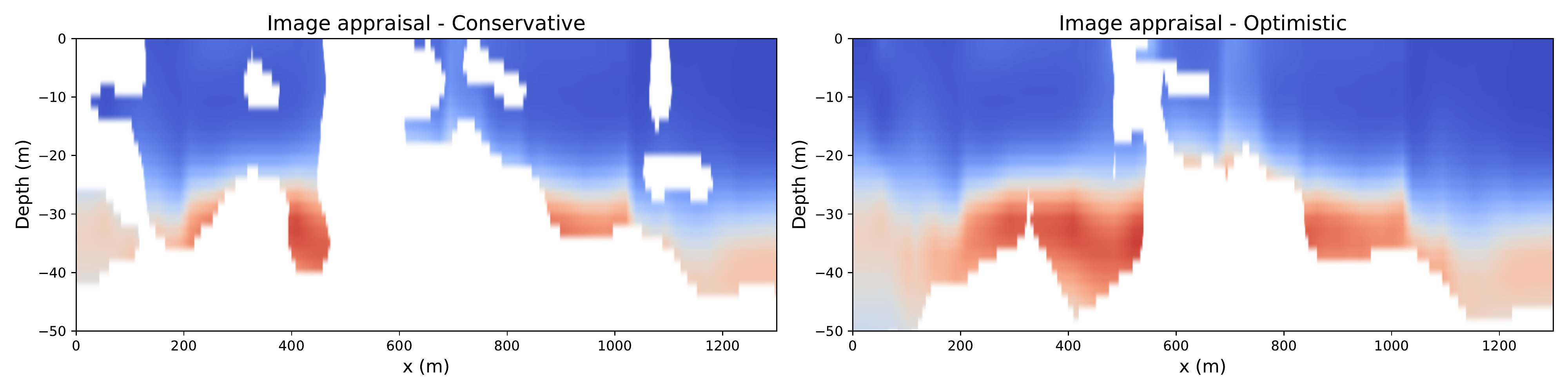} 
	
	\caption{Image appraisal on real field data \add{with \textit{perfect} forward modelling approach. The individual noise-weighted errors in A. include observed noise.}}
	\label{fig:field_perfect}
\end{figure}

\begin{figure}
	\footnotesize 	A.  \hspace{0.5\textwidth} 	\footnotesize 	 B.\\
	\includegraphics[width=0.49\textwidth]{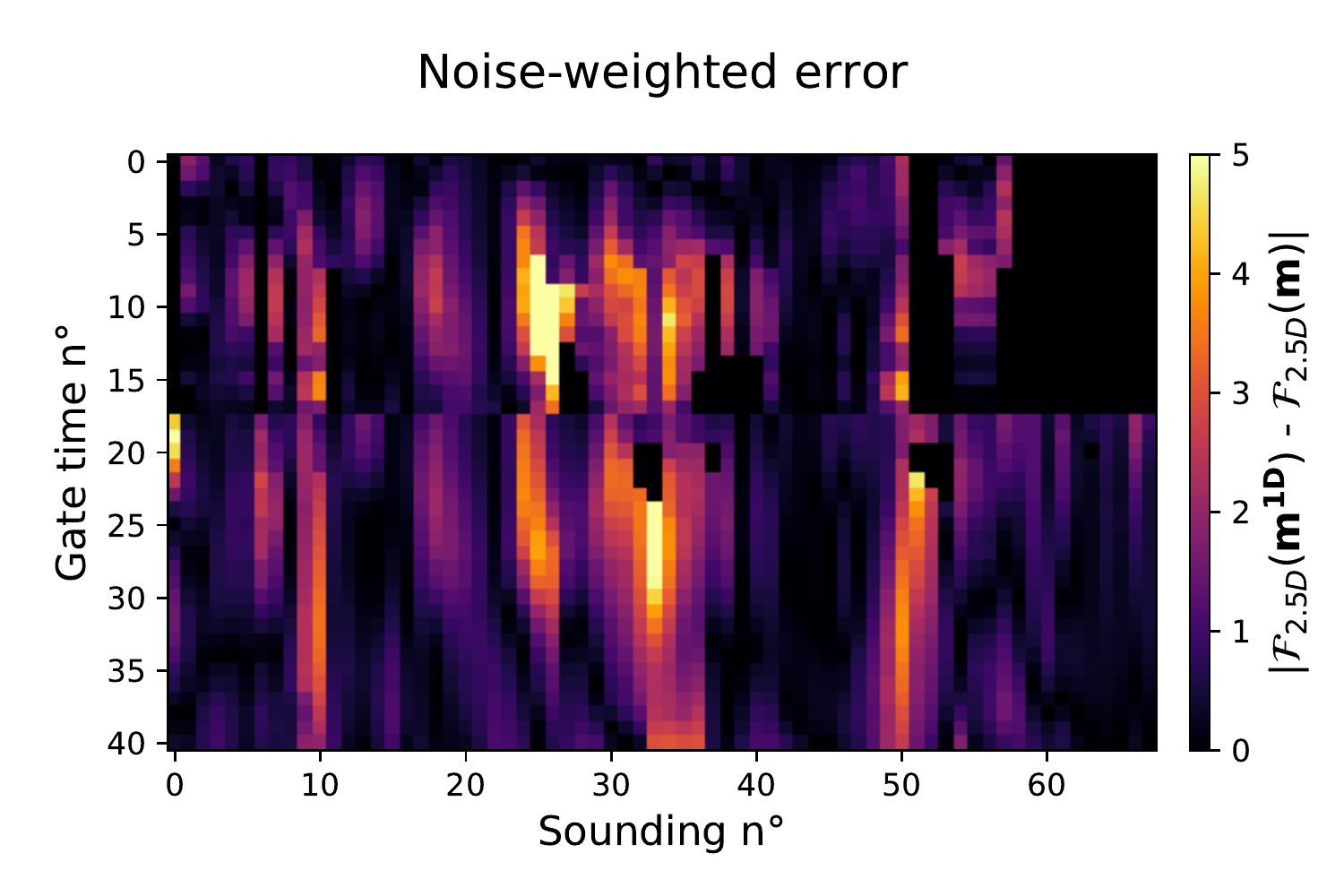} 
	\includegraphics[width=0.49\textwidth]{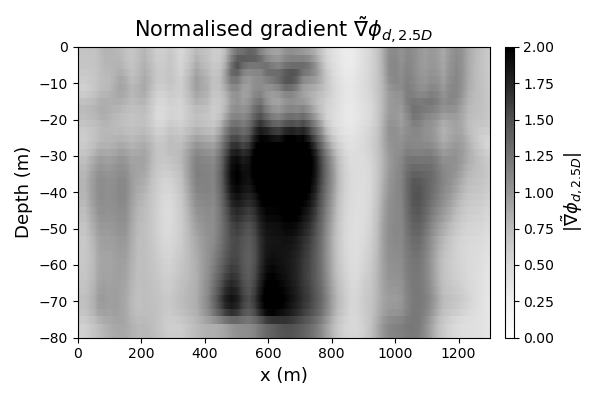} 
	\footnotesize 	C.  \hspace{0.5\textwidth} 	\footnotesize 	 D.\\
	\includegraphics[width=1\textwidth]{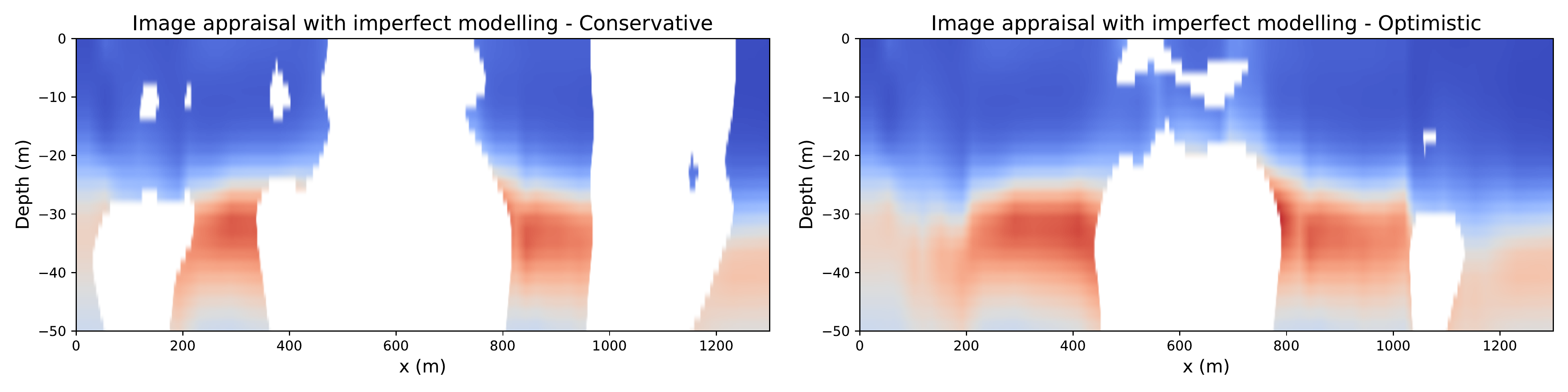} 
	
	\caption{Image appraisal on real field data with \textit{imperfect} forward modelling approach. \add{The individual noise-weighted errors in A. do not include observed noise.}}
	\label{fig:field_imperfect}
\end{figure}

The normalised gradient $\tilde{\nabla} \phi_{d, 2.5D} $ for both approaches is shown in Figures \ref{fig:field_perfect} and \ref{fig:field_imperfect}. In this case one can see the effect of working with the observed data $\vb{d}^{\text{obs}}$  and the medium fidelity data $\mathcal{F}_{\text{2.5D}}  (\vb{m}^{\text{1D}})$ well. The recovered model in Figure \ref{fig:field_recovered} generally fits the late gate times less well (the broad layer around gate time n$^\circ$30 in Figure \ref{fig:field_perfect}A), creating  a relatively present dark layer in the normalized gradient from a depth of 30 m onwards, not necessarily due to multidimensionality issues. The perfect modelling (Figure \ref{fig:field_perfect}) gives a more general picture of which areas of the recovered model do not fit the observed data well\add{, whatever the reason (multidimensionality, misfit related to noise)}. The optimistic image appraisal in Figure \ref{fig:field_perfect}D shows that especially the fresh-saltwater interface around 600-800 m should not be interpreted quantitatively. This is also the conclusion from the image appraisal with imperfect modelling in Figure \ref{fig:field_imperfect}D. Here, the observed data is no longer used and one only gets a picture of areas that do not fit the multidimensionality well (the band around gate time n$^\circ$30 is missing in Figure \ref{fig:field_imperfect}A, while poorly fitting data points are now focused near specific soundings). Our image appraisal analysis teaches one that multidimensional inversion would be advised in the part from 400 to 800 m. \\

Also note that the identified areas are not only zones with sharp conductivity variations. One could argue that the user who knows the 1D approximation would be careful to interpret in such zone, but the results illustrate that this is not sufficient. On the other hand, the tool does not tell if a 2D inversion would result in a significantly different 2D/3D inversion. It primarily indicates zones that might be sensitive to multidimensionality effects.


\section{Conclusions}
We have proposed a computationally inexpensive image appraisal tool for AEM inversion. It enables to assess an inversion model obtained with a low fidelity (approximate) forward model for areas that are not fitting the true multidimensionality of the observed data, because it deviates from the 1D assumption. Adding this step to any quasi-2D or -3D method prevents from quantitatively interpreting the problematic areas in the inversion model. \add{If sufficient computational resources are available, the uncertain zones in the recovered model can be reinterpreted with 3D inversion, instead of performing 3D inversion on the whole dataset.} Furthermore, a forced modelling approach allows to use computationally \change{cheaper}{less demanding} simulations on an imperfect discretization\add{ to identify zones for which multidimensionality likely plays a role}.\\

Thinking about general applications, cf. our method, the specific context of the reader's research will determine which of the two approaches is more desirable. If a HPC infrastructure is available, the perfect modelling will generate more general appraisal images (locating areas which are poorly fitting the observed data). When limited computational resources are available, the forced modelling error approach is more feasible (only requiring a fraction of the perfect modelling resources), focusing on multidimensionality issues only.

\section*{acknowledgments}
The research leading to these results has received funding from FWO (Fund for Scientific Research, Flanders, grant 1113020N) and the Flemish Institute for the Sea (VLIZ) Brilliant Marine Research Idea 2022.. The resources and services used in this work were provided by the VSC (Flemish Supercomputer Center), funded by the Research Foundation - Flanders (FWO) and the Flemish Government.

\section*{Data Availability}
The field data from \citet{vlaanderentopsoil} used for the analysis in Section \ref{sec:field} is available at Zenodo \citep{dataset_306025} via \url{https://doi.org/10.5281/zenodo.7015419} with CC-BY 4.0 license. Scripts that control the above software and produce the appraisal results \citep{AEM_software} are available from Zenodo via \url{https://doi.org/10.5281/zenodo.7015876} with the CC-BY 4.0 license and developed openly at \url{https://github.com/WouterDls/AEM_appraisal}. 

\appendix 
\section{Practical Guide for Constructing a Medium/High-Fidelity Mesh}
\label{sec:app}
\add{
	The multidimensional data is simulated via the finite volume method, a numerical discretization technique for representing and solving partial differential equations (here: Maxwell equations) in the form of algebraic equations. SimPEG} \citep{cockett2015simpeg, heagy2017framework}\add{ is used in this work, which is is an open-source Python package for 3D simulations on a mesh (and other functionalities, such as inversion). As we rely on the moving footprint approach }\citep{cox20103d}\add{, we focus on a suitable mesh that we can use for each sounding (and thus no multiple sources and receivers on one larger mesh). A suitable mesh has smaller cell sizes in crucial areas where the physical quantities significantly vary, e.g. right below the surface. We are working on time-dependent problems, therefore, smaller time steps have to be considered when the physical quantities are changing rapidly. This, in turn, depends on the electrical conductivity (the more resistive, the faster the currents dissipate both downwards and laterally, similar to smoke rings). A good spatial and temporal discretization should be small at crucial stages of the underlying physics to ensure good accuracy, but should also balance the computational burden.\\}

\begin{figure}
	\centering
	\includegraphics[width=0.49\textwidth]{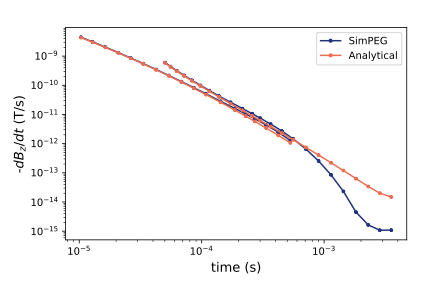} 
	\caption{When comparing the generated data via simulation software with an exact forward model, one can get an indication how to improve the mesh. In this case, where the late time channels break down, the mesh size should be increased.}
	\label{fig:bad_sounding}
	\note{This figure is added w.r.t. the original submission.}
\end{figure}

\add{To construct a mesh, we have used the following strategy: it is a rule of thumb that for time-domain problems, the appropriate cell sizes can be determined from the expected diffusion distance $d$ }\citep{ward1988electromagnetica}\add{, i.e.,}
\begin{equation}
	d = \sqrt{\frac{2t}{\mu\sigma}},
\end{equation}	
\add{where the relation with conductivity is apparent. Note that this expression only holds for homogeneous halfspaces. From this quantity, we determine that the smallest cell size should be no larger than 10\% of the smallest diffusion distance and the thicknesses of the padding should be at least 3 times the maximum diffusion distance. Depending on the specifics of the survey set-up and the expected conductivities, the simulated data for a homogeneous halfspace can be compared to an exact semi-analytical forward model }\citep{hunziker2015electromagnetic}\add{. For example, for a 10 mS/m halfspace, we could get the result from Figure }\ref{fig:bad_sounding}\add{. From a first look, there is a good correspondence for the LM and the early times of the HM. Clearly, at later times, there is a discrepancy. We exploit our understanding of the underlying physics to resolve this issue. The breaking down at late times suggests that the physical dimensions of the mesh are too small (the far away dissipated currents `do not fit' on the mesh). For resistive media, the currents dissipate more quickly and consequently a larger mesh is required. For a homogeneous halfspace of 100 mS/m, the mesh would have been suitable.\\}

\begin{figure}
	\footnotesize 	A.  \hspace{0.5\textwidth} 	\footnotesize 	 B.\\
	\includegraphics[width=0.49\textwidth]{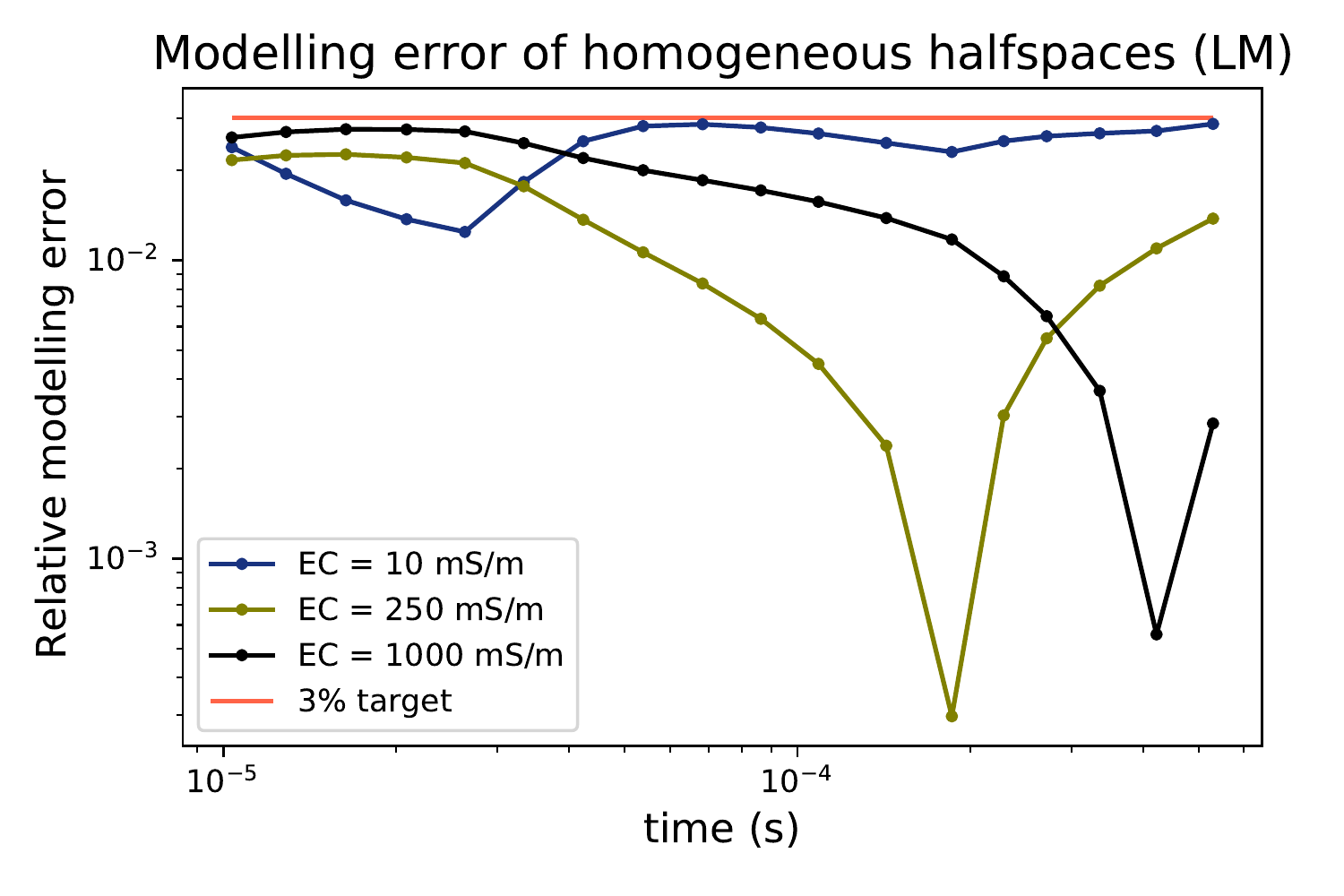} 
	\includegraphics[width=0.49\textwidth]{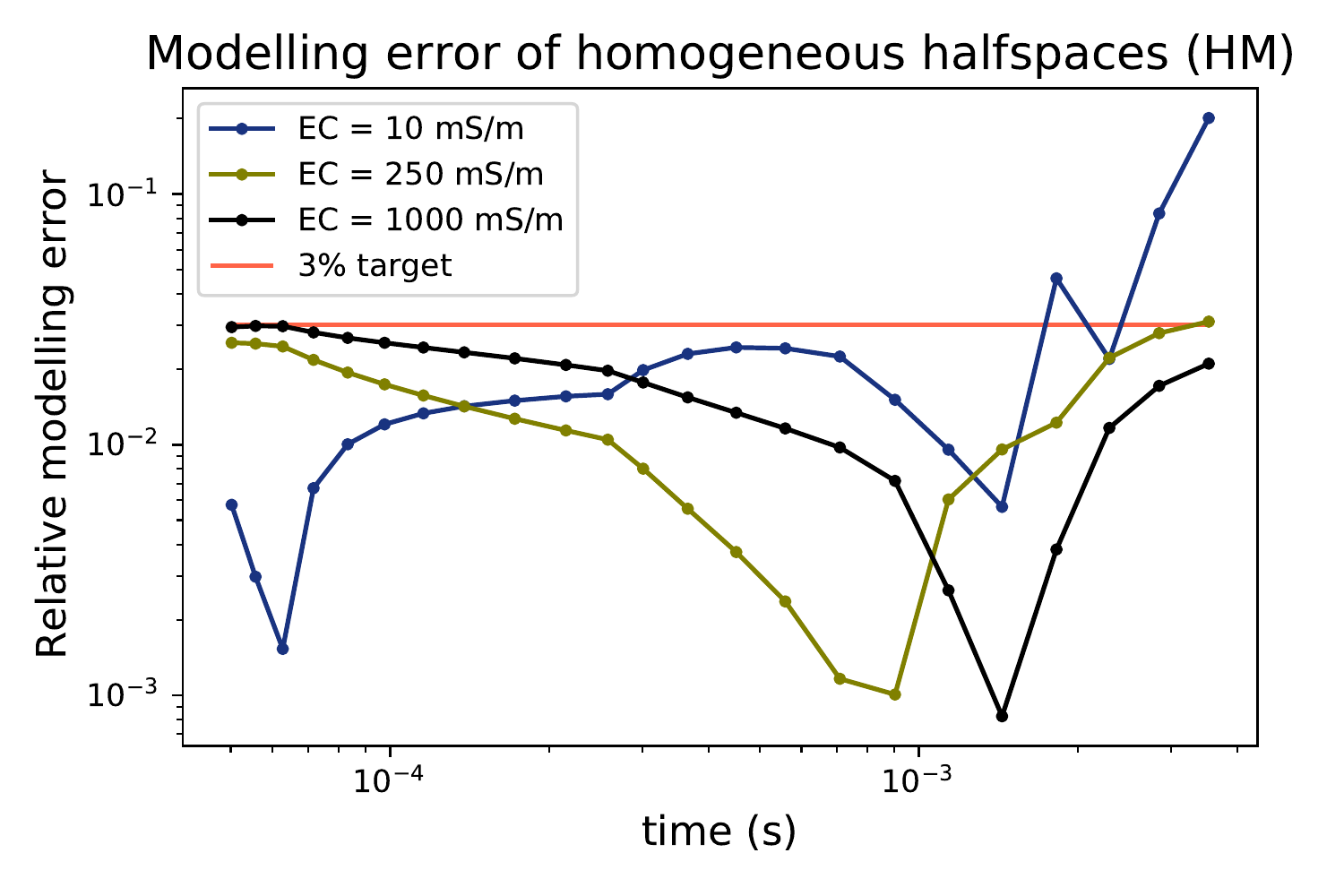} 
	\caption{The relative modelling error of a homogeneous halfspace for the dual moment SkyTEM set-up, compared to the semi-analytical forward model.}
	\label{fig:mesh_targets}
	\note{This figure is added w.r.t. the original submission.}
\end{figure}

\add{A discrepancy at early time gates indicates that the smallest cell sizes are too large. Also, a smaller time stepping could solve the problem. It is a process of trial and error, balancing speed and accuracy.\\
	
	For our work, we have set the target of the modelling error to max. 3\%, a typical measurement noise level. This target has to hold for electrical conductivities ranging between 10 mS/m and 1000 mS/m. The relative modelling error per time channel is shown in Figure }\ref{fig:mesh_targets}\add{A and  }\ref{fig:mesh_targets}\add{B. For the electrical conductivities of  10 mS/m, we abandon the target for later times in the HM, as we will never encouter such a low conductivity for the whole halfspace (the eddy currents will reach a higher conductive area faster than the late times channels, and for higher conductive halfspaces, the accuracy is higher). We also abandon the requirement for the last time channel for all conductivities, as we believe that the additional accuracy is not worth the additional extra computational cost. This is indeed not an issue, because the last time channel is typically close to the noise floor, for which the inverse problem already allows for a larger data misfit.\\
	
	The above rationale was used to construct the high-fidelity mesh for this work. A suitable mesh thus depends on the context of the problem, where setting and loosening the targets should be carefully considered, as well as the computational burden. For the MF, the above requirements do not hold. The mesh is spatially too small (this can be clearly seen in Figure }\ref{fig:synthetic_imperfect}\add{B, where the last time channel has a large discrepancy with the observed data) and the time-stepping is not optimized. The main requirement is always the computational practicality.\\
	
	For constructing the MF mesh, we suggest to start from an HF mesh. For example, by doubling the smallest cell size of the HF mesh and halving the mesh size. For the time stepping, one allows larger time steps at earlier times. One keeps adjusting those parameters until the computation time is acceptable, the user again decides on the balance between computational burden and accuracy. The generated data and gradient will be highly inaccurate, but this poses no problems for the identification of multidimensionality issues: the generated data is solely compared with other data from this MF mesh (with similar modelling errors) and the gradient will still indicate the relevant area in the inversion mesh leading to the specific data (see e.g.,} \citet{zhang20213d}).\\

\begin{figure}[h!]
	\footnotesize 	A.  \hspace{0.5\textwidth} 	\footnotesize 	 B.\\
	\includegraphics[width=0.49\textwidth]{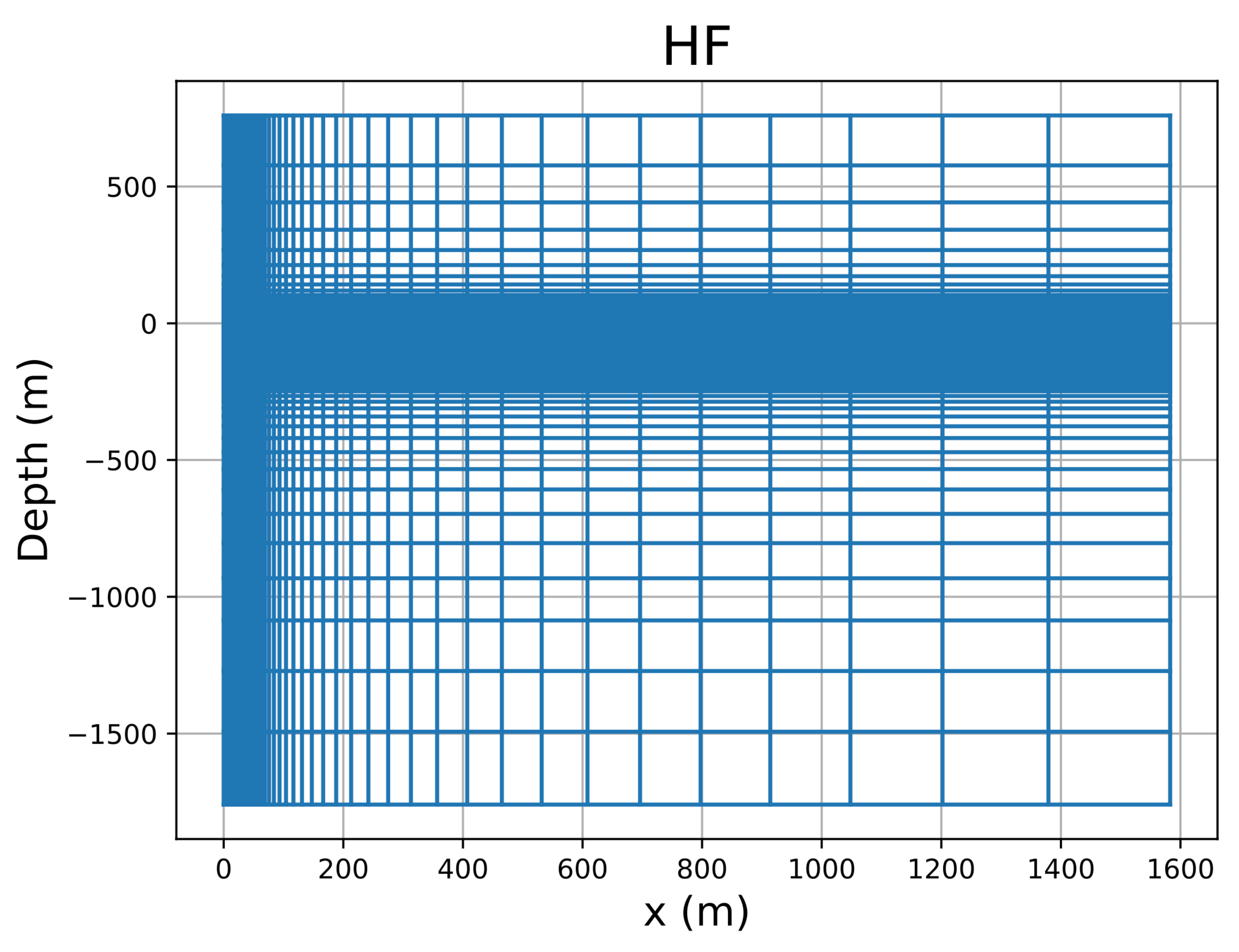} 
	\includegraphics[width=0.49\textwidth]{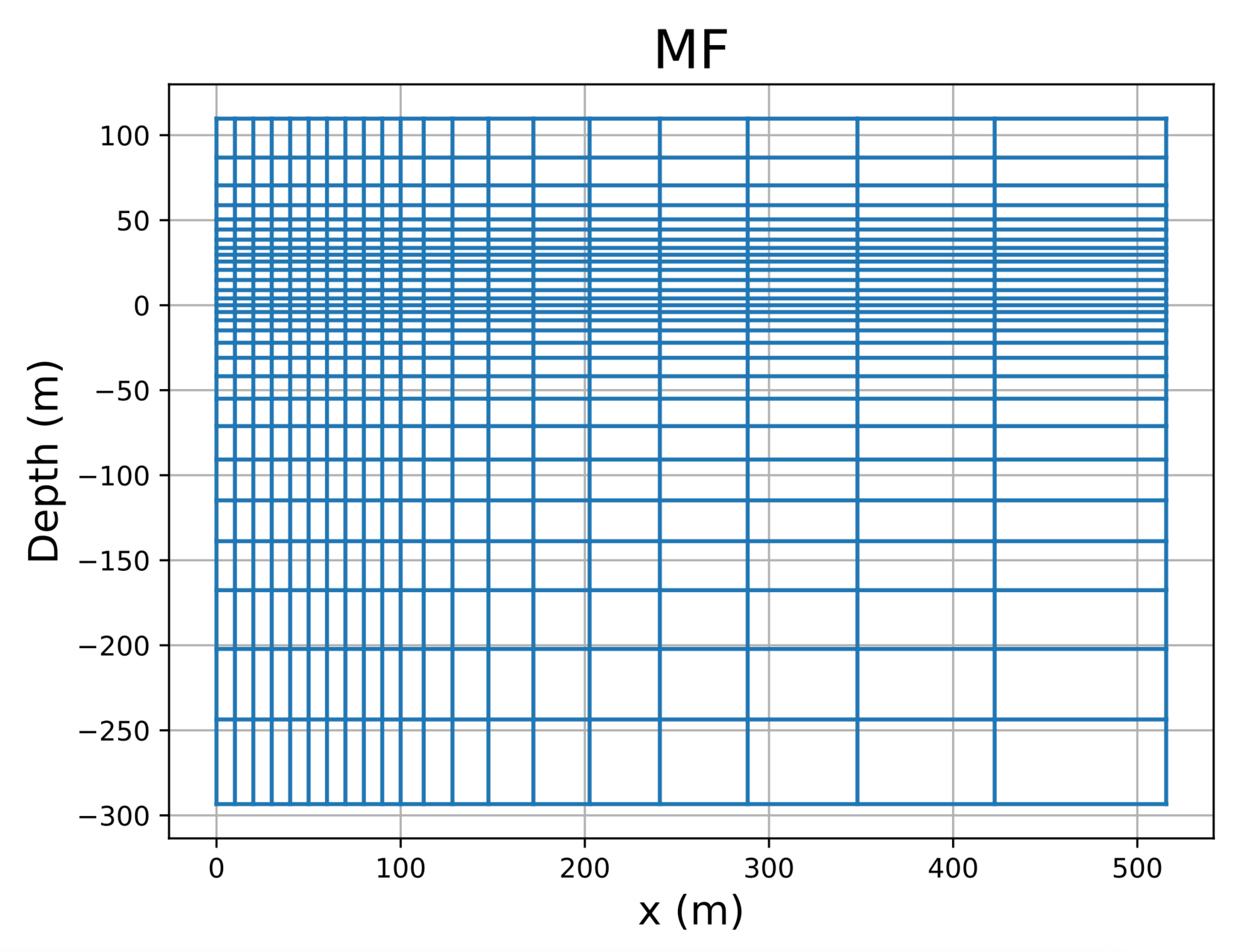} 
	\caption{Cross-section of the 3D meshes used in this work, indicating regions with higher cell densities. A. High-Fidelity (HF) mesh with 1 284 795 cells. B. Medium-Fidelity (MF) mesh with 12 760 cells.}
	\label{fig:mfhf}
	\note{This figure is added w.r.t. the original submission.}
\end{figure}

\add{In this work, we have reduced the HF mesh from Figure }\ref{fig:mfhf}A\add{ with 1 284 795 cells to a MF mesh with 12 760 cells in Figure }\ref{fig:mfhf}\add{B. The smallest cell size is 0.05-by-0.5 and 4-by-10 for the HF and MF mesh, respectively. The spatial dimensions are also reduced: the furthest point on the HF mesh is at 1500 m from the origin, while it is at 675 in the MF mesh. In the HF mesh, 700 time steps are considered, while only 90 time steps are considered in the MF mesh. The specific details of both meshes can be found in }\citet{AEM_software} and \citet{ML_paper}.

\newpage 
\bibliographystyle{apalike}

\label{lastpage}

\end{document}